\documentclass[12pt]{article}

\usepackage[utf8]{inputenc}
\usepackage[T1]{fontenc}
\usepackage{charter}
\usepackage{fullpage}
\usepackage{hyperref}
\usepackage{amssymb}
\usepackage{graphicx}
\usepackage{lipsum}
\usepackage{hyperref}
\usepackage{upgreek}
\usepackage{booktabs}
\hypersetup{hidelinks}
\usepackage[colorinlistoftodos]{todonotes}
\setuptodonotes{inline}
\usepackage[hybrid]{markdown}
\usepackage[total={6.5in,9in},top=1in,headsep=0.1in,headheight=1in]{geometry}
\usepackage{ulem}
\usepackage{cancel}

\usepackage[backend=biber,style=numeric,citestyle=numeric-comp,giveninits=true,isbn=false,date=year,sorting=none]{biblatex}
\renewbibmacro{in:}{%
  \ifentrytype{article}{}{\printtext{\bibstring{in}\intitlepunct}}}
\AtEveryBibitem{%
  \clearfield{number}}
\DeclareSourcemap{
  \maps[datatype=bibtex]{
    \map[overwrite]{
      \step[fieldsource=doi, final]
      \step[fieldset=url, null]
      \step[fieldset=eprint, null]
    }  
  }
}
\DeclareSourcemap{
  \maps[datatype=bibtex]{
    \map[overwrite]{
      \step[fieldsource=eprint, final]
      \step[fieldset=url, null]
    }  
  }
}
\usepackage[ruled,vlined]{algorithm2e}
\usepackage{subfigure}
\usepackage{wrapfig}
\usepackage{booktabs}
\usepackage{multirow}
\usepackage{makecell}
\usepackage{enumitem}

\usepackage{ulem}
\usepackage{bm}

\usepackage[capitalize,nameinlink]{cleveref}
\title{Machine Learning Techniques applied to an Imperfect HEP Dataset: Limitations and Optimization} 
\title{Position Reconstruction using machine learning techniques in cryogenic solid state detectors for dark matter }
\title{Strategies for Machine Learning applied to noisy HEP datasets: A SuperCDMS example}
\title{Strategies for Machine Learning Applied to Noisy HEP Datasets: Modular Solid State Detectors from SuperCDMS}

\date{}

\usepackage{authblk}

\author[1]{P.~B.~Cushman}
\author[1]{M.~C.~Fritts}
\author[2]{A.~D.~Chambers}
\author[3]{A. Roy}
\author[4]{T.~Li}

\affil[1]{School of Physics and Astronomy, University of Minnesota, Minneapolis, MN 55455, USA}
\affil[2]{Department of Physics, Massachusetts Institute of Technology, Cambridge, MA 02139, USA}
\affil[3]{National Center for Supercomputing Applications, University of Illinois at Urbana-Champaign, Urbana, IL 61801, USA}
\affil[4]{Department of Computer Science and Engineering, University of Minnesota, Minneapolis, MN 55455, USA}

\begin{document}

\maketitle
\begin{abstract}
Background reduction in the SuperCDMS dark matter experiment depends on removing surface events within individual detectors by identifying the location of each incident particle interaction.  Position reconstruction is achieved by combining pulse shape information over multiple phonon channels, a task well-suited to machine learning techniques. Data from an Am-241 scan of a SuperCDMS SNOLAB detector was used to study a selection of statistical approaches, including linear regression, artificial neural networks, and symbolic regression. Our results showed that simpler linear regression models were better able than artificial neural networks to generalize on such a noisy and minimal data set, but there are indications that certain architectures and training configurations can counter overfitting tendencies.  This study will be repeated on a more complete SuperCDMS data set (in progress) to explore the interplay between data quality and the application of neural networks. 

\end{abstract}



\section{Introduction}
\label{sec:intro}
The SuperCDMS experiment~\cite{supercdms} is a direct dark matter search performed with modular cryogenic solid-state detectors.  Each detector is an ultrapure disk of germanium or silicon, roughly the size of a hockey puck. The detectors are stacked into towers of 6 each and the towers are  are operated at cryogenic temperatures within a shielded cryostat in an underground lab.   
Interactions with incident particles releases ionization charges and athermal phonons, which can be collected on the top and bottom faces of each detector by thousands of sensors organized into multiple channels.  Signal partition among the channels and the individual pulse shapes themselves provide information on the location of the interaction. 

The internal physics of the phonon and charge transport within the crystal is poorly understood, making it difficult to model the resulting pulse shapes and shared channel behavior from first principles.  Therefore, it is useful to explore machine learning (ML) techniques, which can search broadly over a complex parameter space to reveal correlations and identify the most salient features. We report on the first study using ML to create position reconstruction algorithms in a prototype SuperCDMS detector illuminated by a in-situ movable radioactive source. 

This project is part of the FAIR4HEP initiative\footnote{\url{https://fair4hep.github.io/}} which uses high-energy physics as a science driver for the development of community-wide FAIR  (\textbf{F}indable, \textbf{A}ccessible, \textbf{I}ntero-perable, and \textbf{R}eusable) frameworks~\cite{wilkinson2016fair} to advance our understanding of AI and provide new insights to apply AI techniques. Thus, while this paper presents our own exploration of the efficacy of AI as applied to a small and inherently noisy data set, the data is accessible~\cite{Data}, the results are preserved in a reproducible fashion in line with recent adaptation of FAIR principles for AI applications~\cite{duarte2023fair, huerta2023fair} and researchers are encouraged to expand on our study\footnote{\url{https://github.com/FAIR-UMN/FAIR-UMN-CDMS/}}. 
A second data set with improved area coverage and higher statistics is in preparation and, in comparison with the current data set, will provide important insights into the limitations and strategies required when ML techniques are applied to increasingly more complex data sets.



\section{The Cryogenic Dark Matter Search}
\label{sec:CDMS}

The nature of dark matter is still an outstanding mystery in particle astrophysics. Its existence is needed to explain the flat rotation curves of spiral galaxies, structure formation and galaxy cluster evolution, and the anisotropy of the cosmic microwave background. These astronomical observations indicate that approximately 85\% of the matter in the universe must be dark matter, but they do not tell us its composition. Direct detection dark matter experiments are facilities built on Earth to explore the interaction of this cosmological dark matter with an instrumented target material, in order to determine its particle properties and how it is related to the Standard Model of Particle Physics. 

The Cryogenic Dark Matter Search (CDMS) is one such experiment, which has been operating since the late 90's, with continued improvements in detector sensitivity and exposure.  The latest installment is called SuperCDMS SNOLAB~\cite{SCDMS2017,SuperCDMS_SNOWMASS22}, which will operate a set of 24 solid state detectors 2 km underground in a mine in Sudbury, Canada.  While going underground filters out many of the background-producing cosmic rays, there is still the environmental background from particles generated by trace radioactivity in the surroundings. One background reduction strategy is to cut out all events near the surface of the detector where the incoming radiogenic particles deposit their energy.  Such a fiducial cut can be performed on an entire detector volume by noble liquid experiments,  but modular detectors need to reject those that occur near the surface of each individual detector, requiring good position information.

Determining the location of the initial interaction is complicated since the released energy is shared by all the sensors, with pulse shapes that depend on the details of the phonon physics and charge transport inside the crystal. SuperCDMS uses signal partition between multiple channels to determine the position of a particle interaction. Traditional methods~\cite{SCDMS2018} include charge-sharing and relative amplitude parameters, but also parameters such as the delay between charge and phonon signals.  The phonon risetime is also very sensitive to surface effects and the ratio of primary to ionization-induced phonons~\cite{CDMSLite}.   While simulations from first principles are being actively pursued~\cite{Kelsey}, machine learning techniques can provide a way to reconstruct the location of a particle interaction within the crystal by training on datasets where the incident particle position is known, e.g.\ by mapping the surface of the detector with a radioactive source. 

\subsection{The SuperCDMS SNOLAB Detectors}



\begin{figure}
   \centering
    \includegraphics[width=0.9\textwidth]{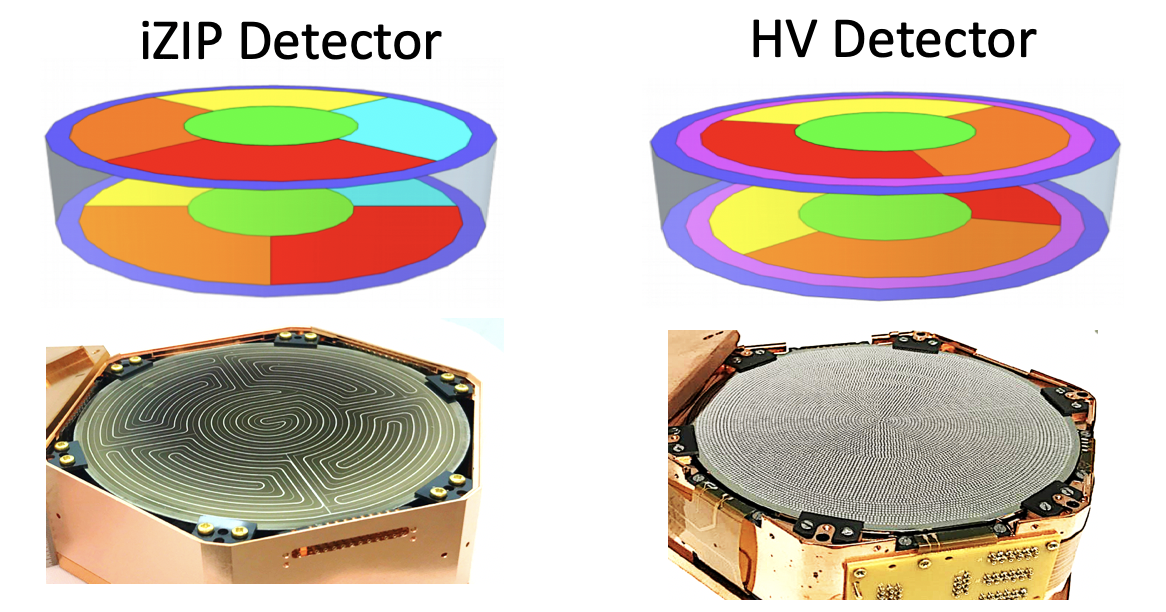}
    \caption{The SuperCDMS detectors come in two sensor arrangements: Interleaved Z-sensitive Ionization \& Phonon (iZIP) and High Voltage (HV). The arrangement of the 12 channels formed by grouping together individual transition edge sensors (TES) are shown above photographs of the detectors. }
    \label{fig:Detectors}
\end{figure}

Each detector is a 10 cm diameter, 3.3 cm thick disk of either silicon or germanium. A particle that interacts with the crystal ionizes the medium, creating charge that can be collected by electrodes on both faces of the detector.  Interactions also produce phonons (quantized vibrations of the crystal lattice) which are read out using a mosaic of transition edge sensors (TES) interleaved with the electrodes. The sensors are connected in parallel within a channel, creating a 6-fold segmentation on each face of the detector. 
Depending on the configuration of the applied sensor array, the detector is defined as an iZIP or an HV (see Figure \ref{fig:Detectors}).  The iZIP runs at low bias voltage and reads out both charge and  phonon signals to achieve discrimination between electron and nuclear recoils~\cite{CDMS2014}, whereas the HV detector  runs at high bias voltage to increase sensitivity to small energy depositions~\cite{CDMS2016}.

Since the HV detector does not have an independent charge read out, we must rely on developing position-dependent fiducial cuts based only on the phonon channels. This drives the channel design, with two outer rings for improved fiducialization. However, a simple veto is not possible since, as explained in the following section, the formation and readout of phonon pulses takes hundreds of microseconds and intersects all the channels. 

\subsection{Pulse formation and Readout}

An incident particle scatters within the crystal and induces vibrations at the position of the recoiling nucleus or electron. These primary phonons expand outward for several microseconds. During this initial stage, the phonon interaction length is very short, so they propagate quasi-diffusively, decaying rapidly to lower energy states. Once the phonons reach energies where their mean free path becomes of the order of the size of the detector, they become ballistic and scatter between the bare crystal surfaces until they encounter a sensor. Upon encountering a sensor, the energy from the primary phonon is collected.  Due to the small proportion of surface area covered by sensors, it can take 50-100 scatters for all the ballistic phonons to be absorbed. Thus, an event caused by a particle interacting in the detector produces a waveform with two distinctive characteristics: a
sharp initial peak which gives localization information and a slow decay which contains energy information.  The overall pulse shape encodes for geometry, surface effects, and bulk properties.  

The Transition Edge Sensors (TES) are thin tungsten wires held at their superconducting transition temperature $(T_c)$. When phonons with sufficient energy interact with the sensor, they will break up a Cooper pair and create quasiparticles. These quasiparticles are progressively trapped in a W/Al overlap region where they scatter, lose energy, and eventually cannot re-enter the aluminum. The energy lost by the phonon in this process is collected by the voltage-biased TES and finally converted to a voltage waveform by a SQUID array. The relaxation of the readout circuit back to baseline contributes to the shape of the falling edge of the waveform. 

Electron-hole pairs are also created at the location of the initial recoil.  A bias voltage is applied across the crystal via electrodes patterned on both faces of the detector.  This produces a uniform electric field within the crystal, which drives the electrons to one surface and the holes to the other. As the electric field accelerates the charge carriers, their energy is transferred to the lattice.  The resulting phonons are called Neganov-Trofimov-Luke (NTL) phonons~\cite{Luke1988,Neganov1985} and they are read out along with the phonons from the initial recoil, thus providing phonon amplification of the ionization signal proportional to the bias voltage.

Compared to the recoil phonons, the NTL phonons are more tightly focused above and below the interaction point when under high bias voltage. This is because they are produced along the path taken by the charge carriers. Since the minimum velocity of the charge carrier required to a create an NTL phonon is the speed of sound, the NTL phonons also form a forward-going cone along the charge carrier paths, analogous to Cerenkov radiation, thus providing another handle on position reconstruction.

\begin{figure}
   \centering
    \includegraphics[width=0.9\textwidth]{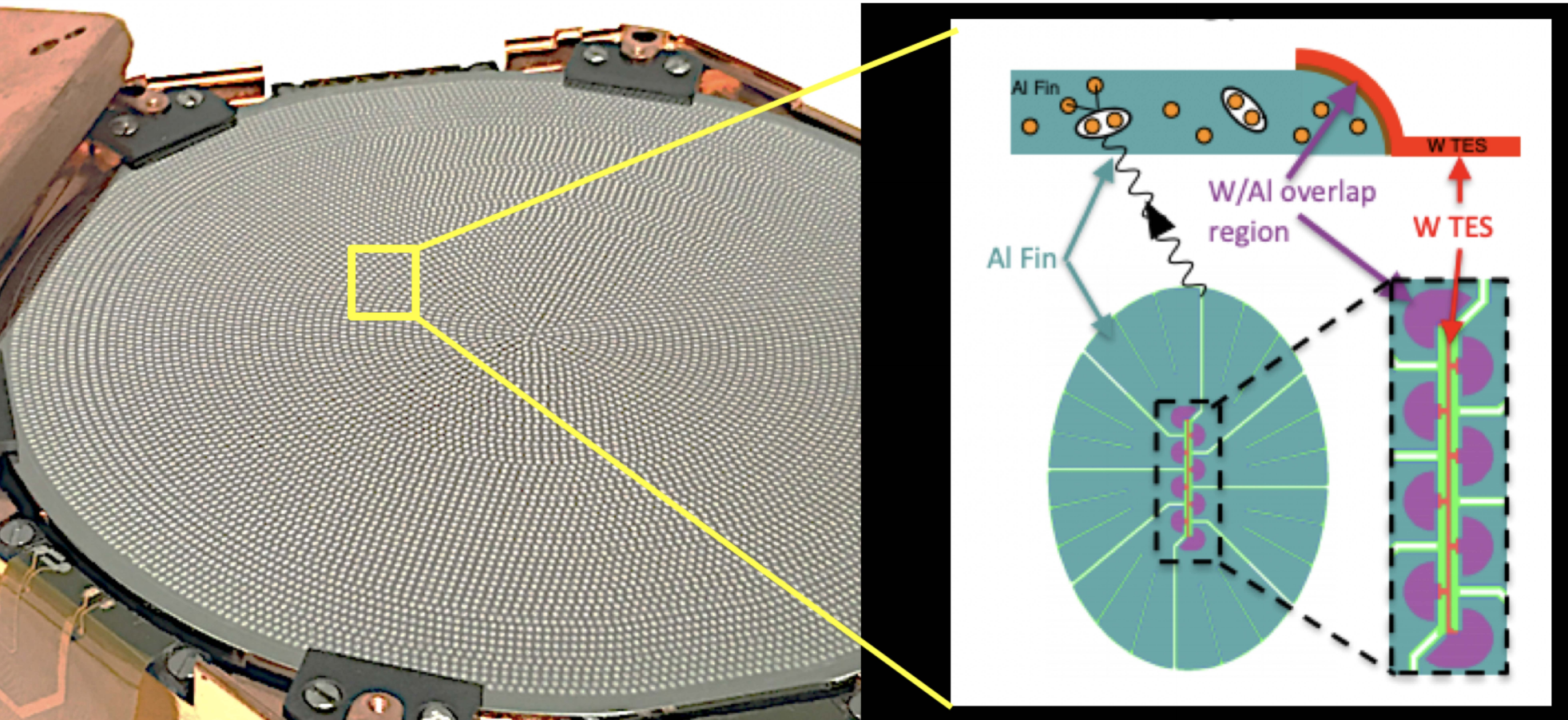}
    \caption{The aluminum collection fins which cover the two faces of the SuperCDMS HV detector can be seen on the left as bright ovals.  On the right panel is a schematic of the quasiparticle trapping process. The quasiparticles are orange dots and the Cooper pairs are shown as double dots. Each oval set of collection fins surrounds a tungsten TES (in red) and its AL/W overlap regions (in purple). }
    \label{fig:sensor}
\end{figure}
\subsection{Position Reconstruction using Channel Parameters}
Simple formulas for x-y position mappings can be constructed based on the relative amounts of phonons reaching neighboring channels (``partition'') and the relative arrival times of the phonons (``delay''). The channels nearer to the interaction will receive phonons sooner and a greater share of them. Each channel is assigned coordinates based on its position. For example, the three HV channels (red, orange, yellow in right panel of Figure~\ref{fig:Detectors}) can be used to define the following mappings based on the pulse amplitudes and start times. The subscripts $amp$ and $start$ represent the amplitude and start time of a pulse, respectively. Based on equivariance of reconstructed $(x,y)$ coordinates under planar rotations and assuming linear dependence between channel observables and reconstructed coordinates, simple projective mappings can be examined-  

\begin{align}
         X_{\text{partition}}  =& \;\frac{E_{\text{amp}}-0.5(C_{\text{amp}}+D_{\text{amp}})}{ C_{\text{amp}}+D_{\text{amp}}+E_{\text{amp}}}\label{eq:xpart}\\
         Y_{\text{partition}}  =& \;\frac{\sqrt{3}/2(C_{\text{amp}}-D_{\text{amp}})}{ C_{\text{amp}}+D_{\text{amp}}+E_{\text{amp}}}\label{eq:ypart}\\
         X_{\text{delay}}  =& \;0.5(C_{\text{start}}+D_{\text{start}})-E_{\text{start}} \label{eq:xdelay}\\
         Y_{\text{delay}}  =&\; \sqrt{3}/2(D_{\text{start}}-C_{\text{start}}) \label{eq:ydelay}
\end{align}
where $C,D,E$ respectively represent the yellow, red, and orange channels in the HV detector module in Figure~\ref{fig:Detectors}, $X$ and $Y$ are the projective coordinate mappings based on pulse amplitudes or start-times. 

\begin{figure}
    \centering
    \includegraphics[width=0.45\textwidth]{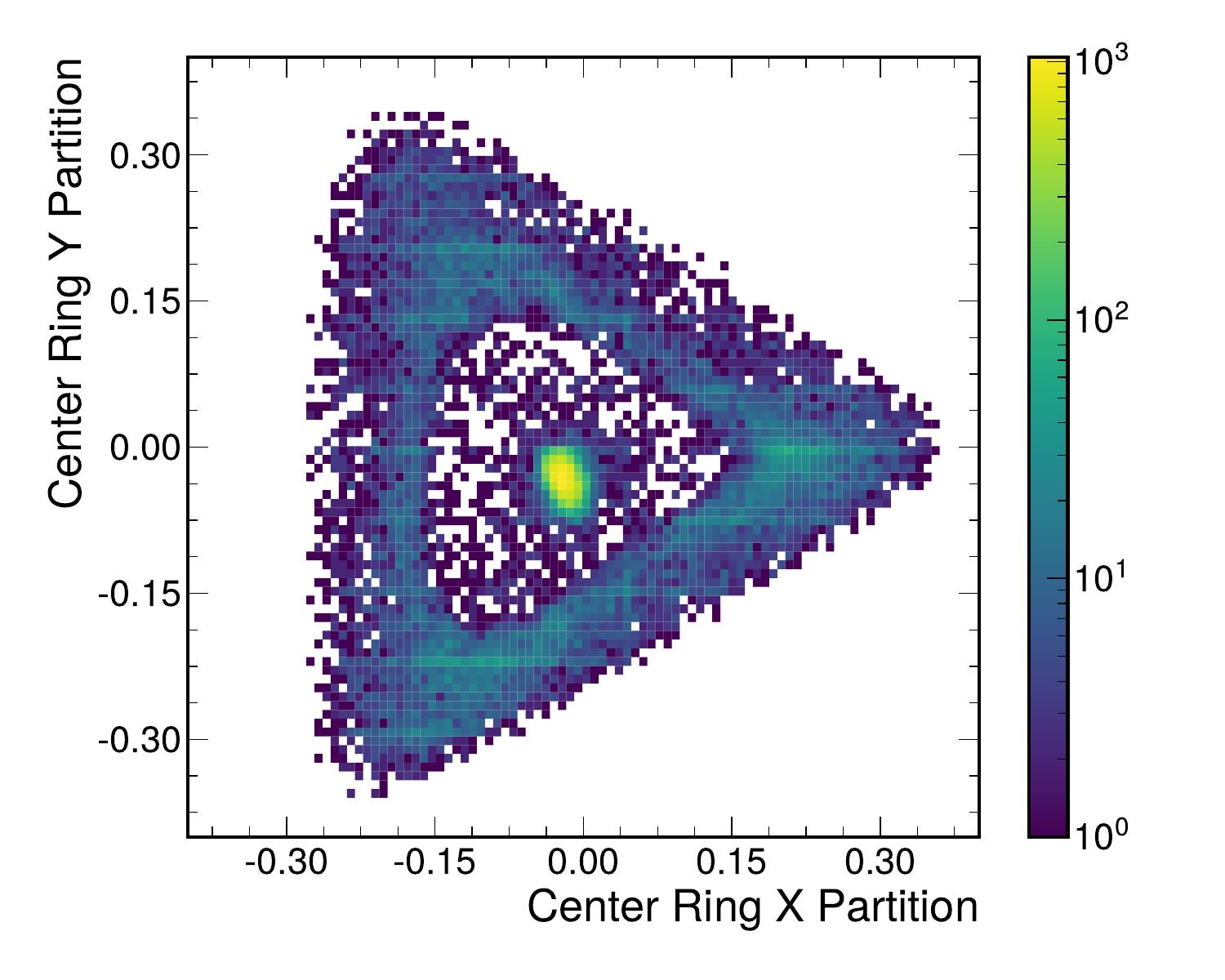}
    \includegraphics[width=0.45\textwidth]{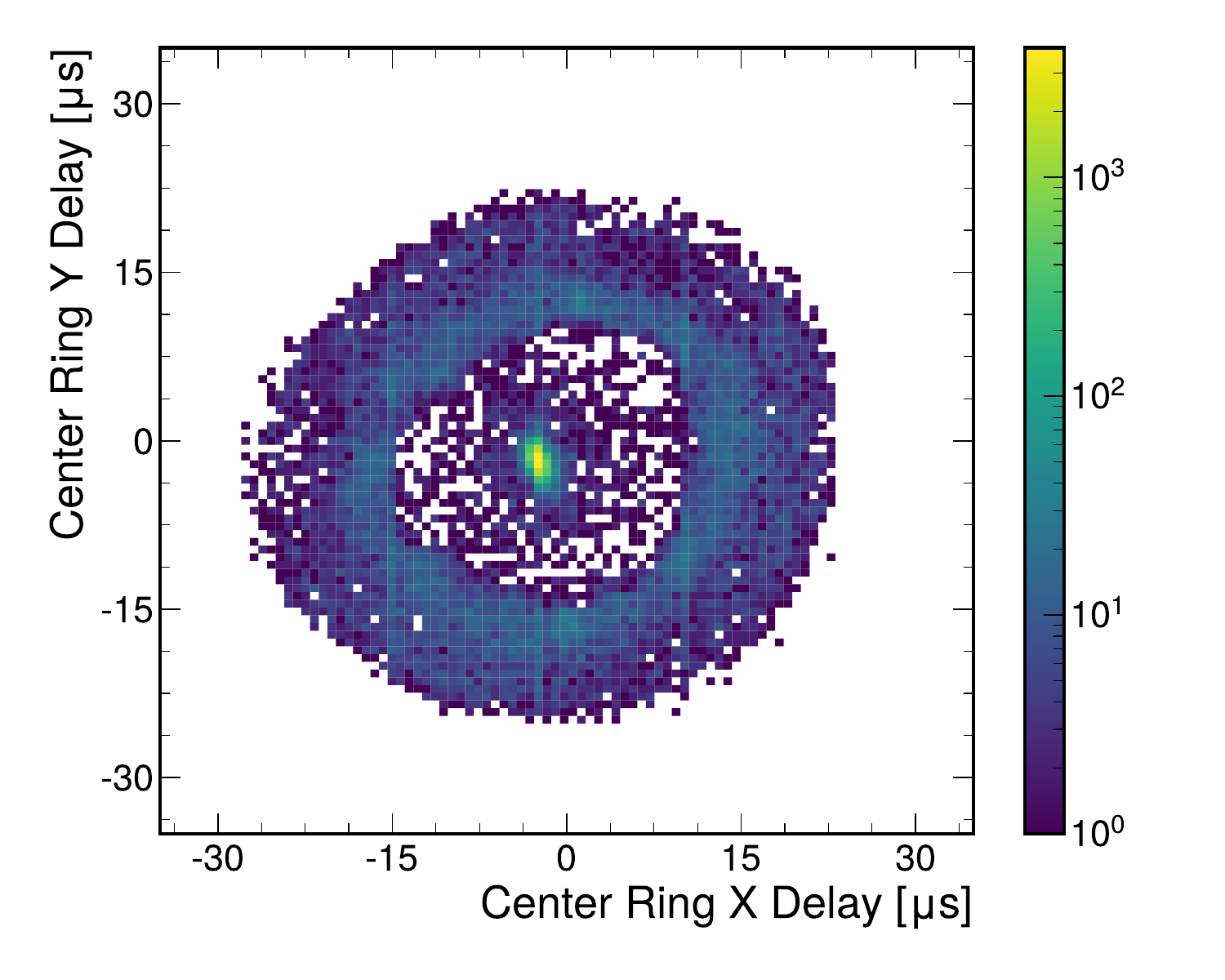}
    \caption{Position mapping using the three inner ring channels on one side of an HV detector.}
    \label{fig:partition_and_delay}
\end{figure}

Figure~\ref{fig:partition_and_delay} plots the resulting positions for a dataset of around 30,000 interactions with deposited energies between 13 and 19~keV. About half of the interactions were produced by a collimated gamma radiation source aimed at the center of the detector.  The spot at the center of both plots is the reconstructed position of these events. Background events from environmental gammas and betas uniformly populate the detector surface. 


If these mappings were representative of true position reconstructions, the figures would show a uniform density. Instead, we see a pattern of non-uniformity in these distributions.  The non-uniformities are different in the two plots, suggesting that partition and delay (amplitude and timing) might provide complementary information in a true position reconstruction problem. These mappings can give some idea of $(x,y)$ positions, but the relationship to true position is difficult to determine. While both figures show distinctive symmetrical properties emerging from the detector geometry, these relationships are also not single-valued, leading to a pattern of degeneracies.  A more sophisticated position reconstruction is desirable - one that potentially combines amplitude and detailed timing information from all channels.

\section{Characteristics of the Dataset}
\label{sec:dataset}

\subsection{Scanning the SuperCDMS HV Detector}
\label{sec:scan}
The University of Minnesota maintains an Oxford Instruments Kelvinox-100 dilution refrigerator characterized by 100~$\mu$W cooling power and a base temperature of $<30$~mK. A prototype SuperCDMS germanium HV detector was installed in the cryostat and operated at a bias of 300 volts. 
In order to inject a position-dependent calibration signal into the detector at cryogenic temperatures, an in-situ source mover was developed~\cite{Mast_2020} by rewinding the coil of a commercial stepper motor with superconducting NbTi wire and using the motor to drive a threaded rod along a linear path.  Mounted on the rod was an Am-241 source in a 1.7~mm thick lead disk. A collimated beam emerged from a 0.46~mm diameter hole 6~mm above the detector surface.  As the source was scanned across the detector, the resulting waveforms were recorded with a 1.25~MHz digitizer, 4096~samples per waveform. 

\begin{figure}
    \centering
    \includegraphics[width=0.36\textwidth]{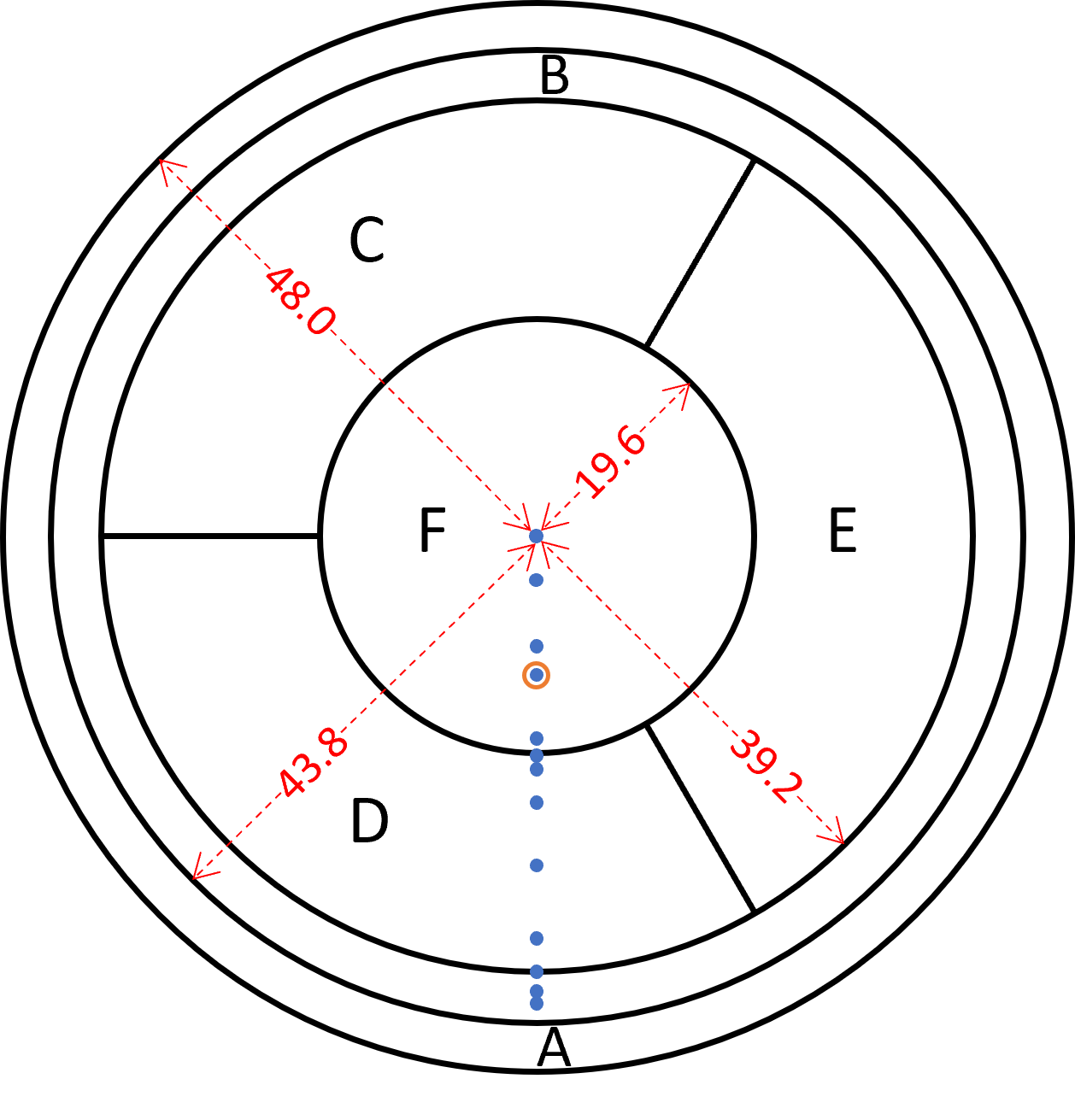}
    \includegraphics[width=0.63\textwidth]{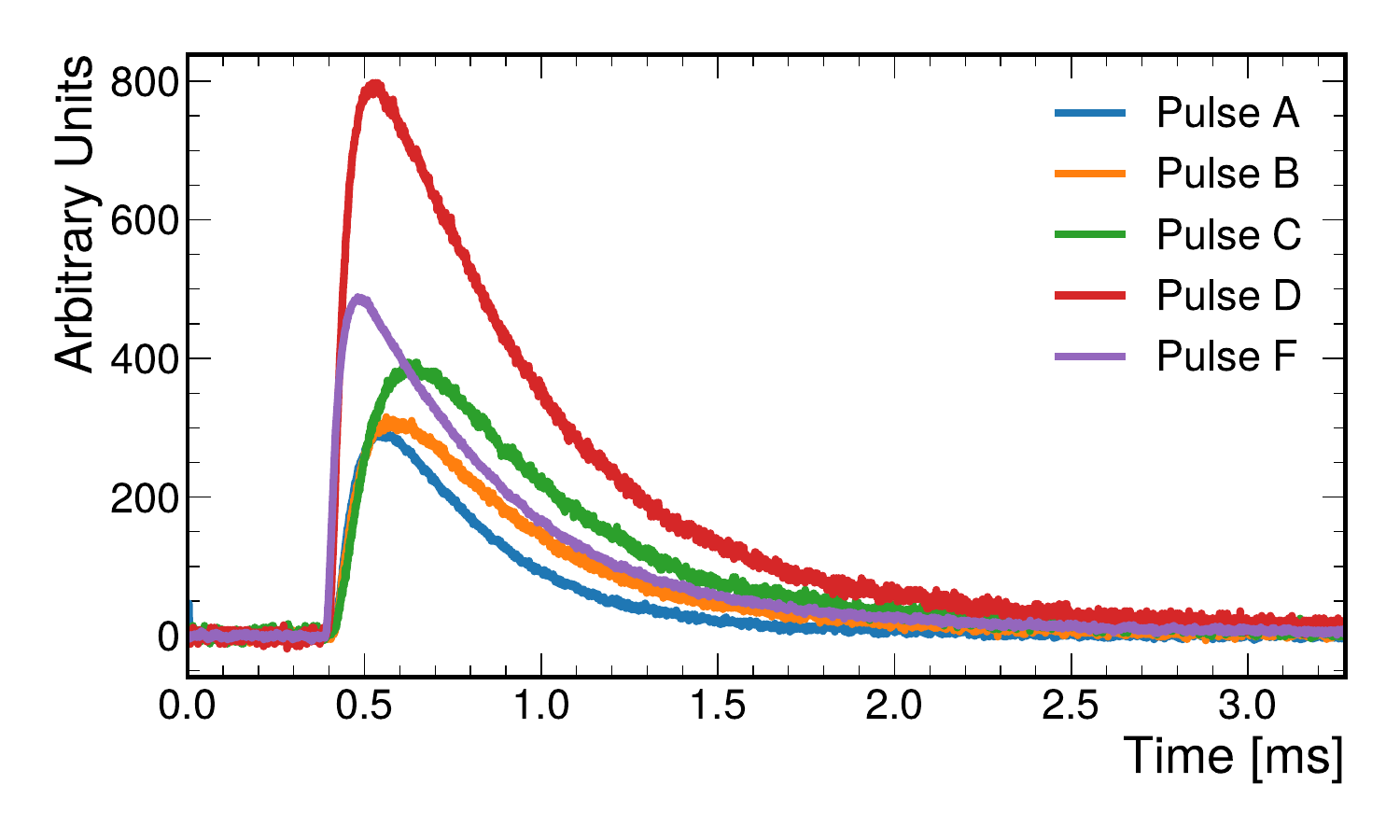}
    \caption{Left: The locations of the Am-241 source during data collection are shown as blue dots superposed on the SuperCDMS HV detector channels (dimensions shown in mm). Right: phonon pulse shapes recorded by individual channels on the top surface of a SuperCDMS SNOLAB HV prototype at 300~V bias, when the Am-241 source is above the circled location. Channel E was not operational for this test. }
    \label{fig:locations}
\end{figure}

The movable radioactive source was used to produce interactions at thirteen different locations on the detector along a radial path from the central axis to close to the detector’s outer edge, as shown in Figure~\ref{fig:locations}. Waveforms from a total of 7151 particle interactions were recorded over these 13 impact locations. The exact locations of the source and the number of interactions are given in~\cref{tab:data_sample_info}.

Figure ~\ref{fig:locations} shows the  waveforms (labeled by their channel) resulting from an interaction with the source at the circled location in the figure.  As the phonons move outward from the interaction point, the closest channels (in this case F and D) are intersected first, forming the sharp risetime and high amplitudes seen in waveforms F and D. All channels collect phonons as the vibration spreads outward, reflecting multiple times on all surfaces, thus forming the long tails. 

\begin{table*}[!htpb]
\caption{The different impact locations and the corresponding number of experiments performed at that location. The entries marked as bold are used as the held-out set, as explained in~\cref{sec:ml-data}, to test the ability of the ML models to generalize for previously unseen data.}
\label{tab:data_sample_info}
\begin{center}
\setlength{\tabcolsep}{5.0mm}{
\begin{tabular}{c c | c c}
\toprule
Location [mm] & Sample number & Location [mm] & Sample number
\\
\midrule
0.0 & 924 & -24.077 & 567\\
-3.969 & 500 & \textbf{-29.500} & \textbf{634}\\
-9.988 & 613 & -36.116 & 560 \\
\textbf{-12.502} & \textbf{395} & -39.400 & 386 \\
-17.992 & 357 & -41.010 & 606 \\
-19.700 & 376 & \textbf{-41.900} & \textbf{486}\\
-21.034 & 747\\
\bottomrule
\end{tabular}

}
\end{center}
\end{table*}

\subsection{Dataset Limitations }
The finite size of the collimator limits the precision with which we know the location of the interaction point.   Monte Carlo simulations indicate that for a given source position the actual interaction location is spread out in $x$ and $y$ with a standard deviation of 0.47~mm. 

Issues with cryogenic stability produced slowly varying and channel-dependent drifts in the signal gains over time. Empirical corrections of the amplitudes were performed to lessen the bias. However, the amplitude information may still include systematic errors since the amplitude correction is applied unevenly across source positions. Note that pulse timing data is unaffected by this problem and should be more reliable.

Further cleaning of the data was performed in order to remove background, isolate the source information and reject bad pulse shapes, as follows: 
\begin{enumerate}
  \item Mismatch between algorithms is used to quantify pulse amplitude. If the integral amplitude fit does not agree with the pulse shape template, the interaction is rejected. This effectively removes pile-up when two interactions occur so close in time that multiple pulses appear in the same waveform. 
  \item Timing outliers due to failure of pulse timing algorithms for distortions or noise in the waveforms are rejected. 
  \item If the fraction of energy in the channel nearest to the radioactive source location is low, it is likely caused by a background interaction in a different channel. These events are rejected. 
  \item The Am-241 source produces characteristic x-ray energies which appear as peaks in plots of distributions of the total energy. We select only interactions in a region of interest around the 14.0 and 17.7 keV peaks.
\end{enumerate}


\subsection{Preparing an ML Dataset}
\label{sec:ml-data}
For each waveform, a set of parameters was extracted from the signals from each of the five sensors. These sets have been compiled into two datasets which we will refer to as the \textit{reduced/simple} (S) and \textit{full/extended} (X) datasets. The \textit{reduced dataset} contains 19 input features which represent information known to be sensitive to interaction location, including the relative timing between pulses in different channels, and features like the pulse shape. The parameters included in the reduced dataset for each channel are:
\begin{itemize}
\item start:
The time at which the pulse rises to 20\% of its peak with respect to Channel A
\item rise:
The time it takes for a pulse to rise from 20\% to 50\% of its peak
\item width:
The width (in seconds) of the pulse at 80\% of the pulse height
\item fall:
The time it takes for a pulse to fall from 40\% to 20\% of its peak
\end{itemize}
The \textit{full dataset} provides 85 input features for each interaction. These include one amplitude parameter and sixteen timing/shape parameters for the waveforms for each of the 5 channels (A, B, C, D, F, since E was inoperable during the run). The amplitude parameter is a measure of the size of each waveform based on comparison to a normalized waveform template and given in arbitrary units. It is labeled by channel, e.g. Bamp. The timing parameters represent the time during the rise and fall of the waveform at which the pulse reaches a given percentage of its maximum height. The parameters are given names such as Cr40 (the 40\%-point for channel C as the waveform is rising) and Ff80 (the 80\%-point for channel F as the waveform is falling). These times are measured with respect to Ar20, the timestamp of the pulse in channel A reaching 20\% of its maximum height. 
Thus, Ar20 is always zero, reducing the number of non-trivial features to 84. In both datasets, the timing information is given in units of micro-seconds ($\mu$s) and the impact location is given in units of millimeters (mm). 

To train the ML models and also test its generalization to new samples with never-seen positions during the process of model learning, we first divide the dataset into two separate subsets:
\begin{itemize}
    \item \textit{Model-learning subset} MLS contains 5636 data samples whose positions are in the set \{0.0, -3.969, -9.988, -17.992, -19.700, -21.034, -24.077, -36.116, -39.400, -41.010\}. The subset is further divided into training and validation sets in a 4:1 ratio. The splitting is done randomly during each instance of model training. Hence, each model is trained multiple times to test its sensitivity to this random split. The training set is used to train the ML models and the validation set is used to  validate the performance of the learned model.
    \item \textit{Held-out subset} HOS contains 1515 data samples whose positions are in the set \{-12.502, -29.500, -41.900\}. This is treated as our test set in which we evaluate the performance of the learned ML model in generalizing to previously unobserved locations on this subset.
\end{itemize}

\begin{figure*}[!htpb]
\centering
\subfigure[]{ {\includegraphics[width=0.48\textwidth]{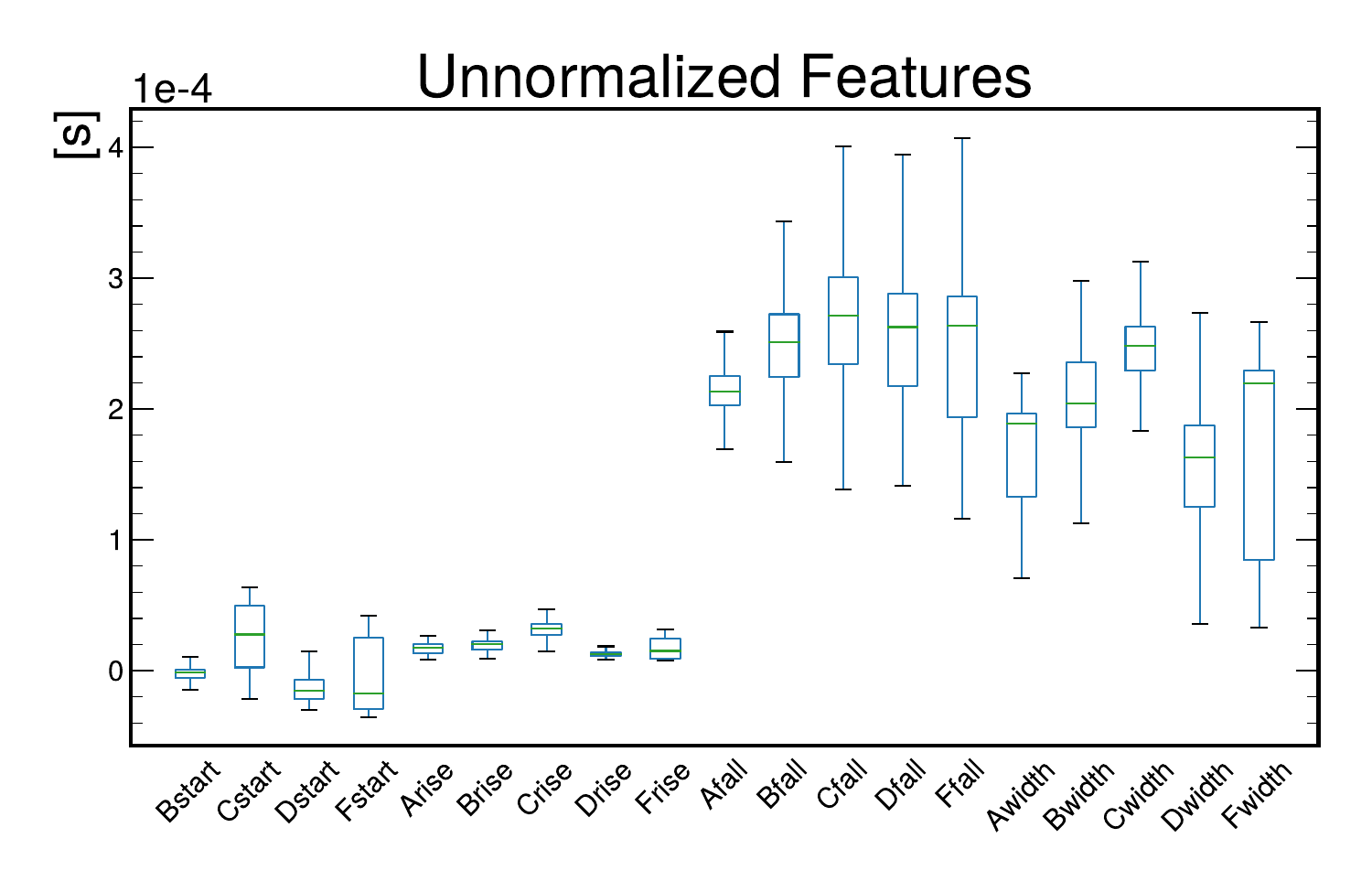}} }
\subfigure[]{ {\includegraphics[width=0.48\textwidth]{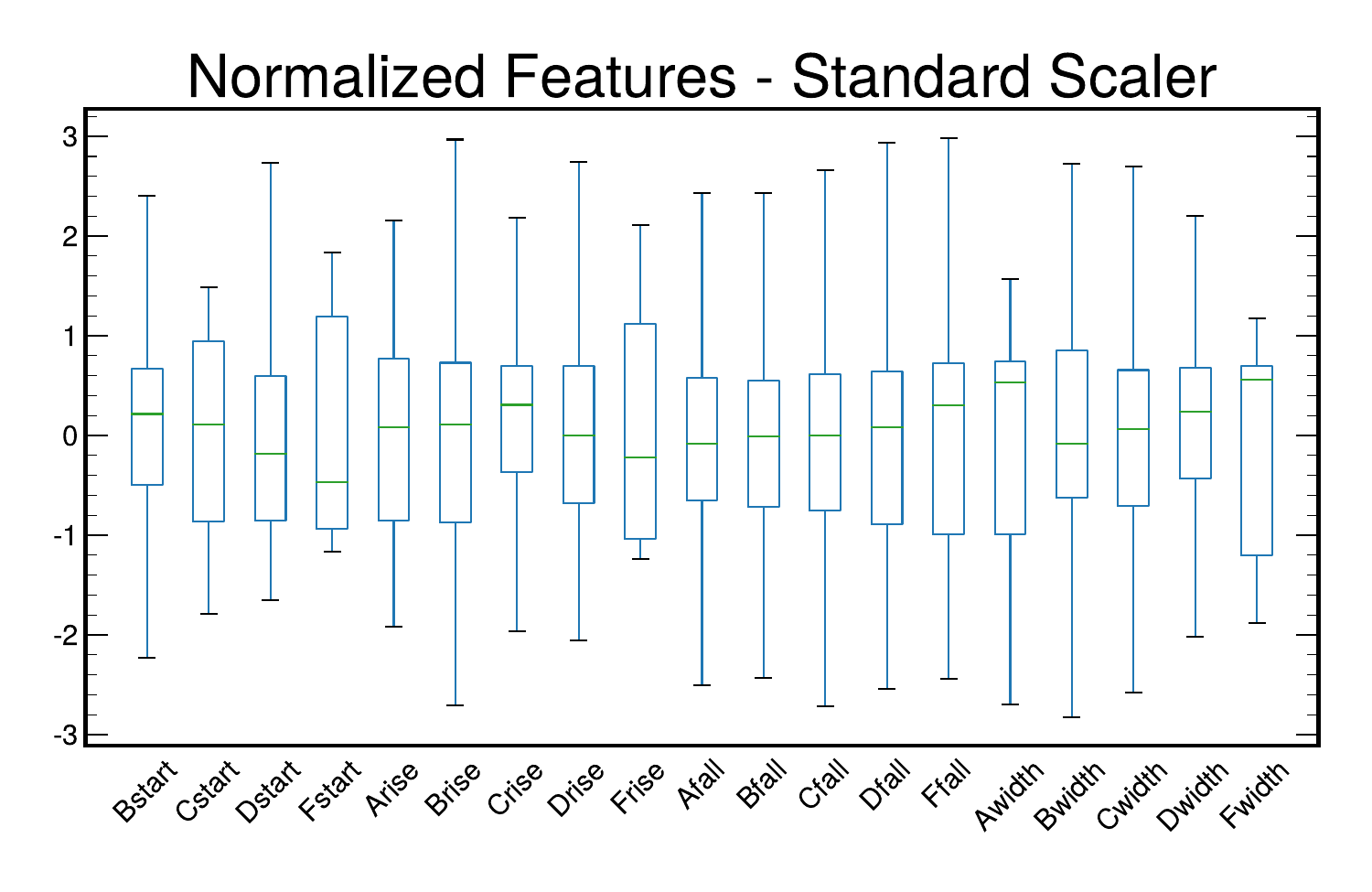}} }
\caption{The numerical range of each feature before normalization (left) and after normalization (right).}
\label{fig:features_normalization}
\end{figure*}

To facilitate the learning process, it is essential to standardize the features. In this work, a standard scaling method is used, where each feature is scaled to have a mean of zero and a standard deviation of one. 
\cref{fig:features_normalization} showcases the feature distributions for the simple dataset before (\cref{fig:features_normalization}(a)) and after (\cref{fig:features_normalization}(b)) standardization. Note that the numerical range widely varies across features,
indicating that it is crucial to standardize the data in order to mitigate the domination of learning by a small subset of features.

\subsection{The Position Reconstruction Problem}

\begin{figure}
    \centering
    \includegraphics[width=0.7\textwidth]{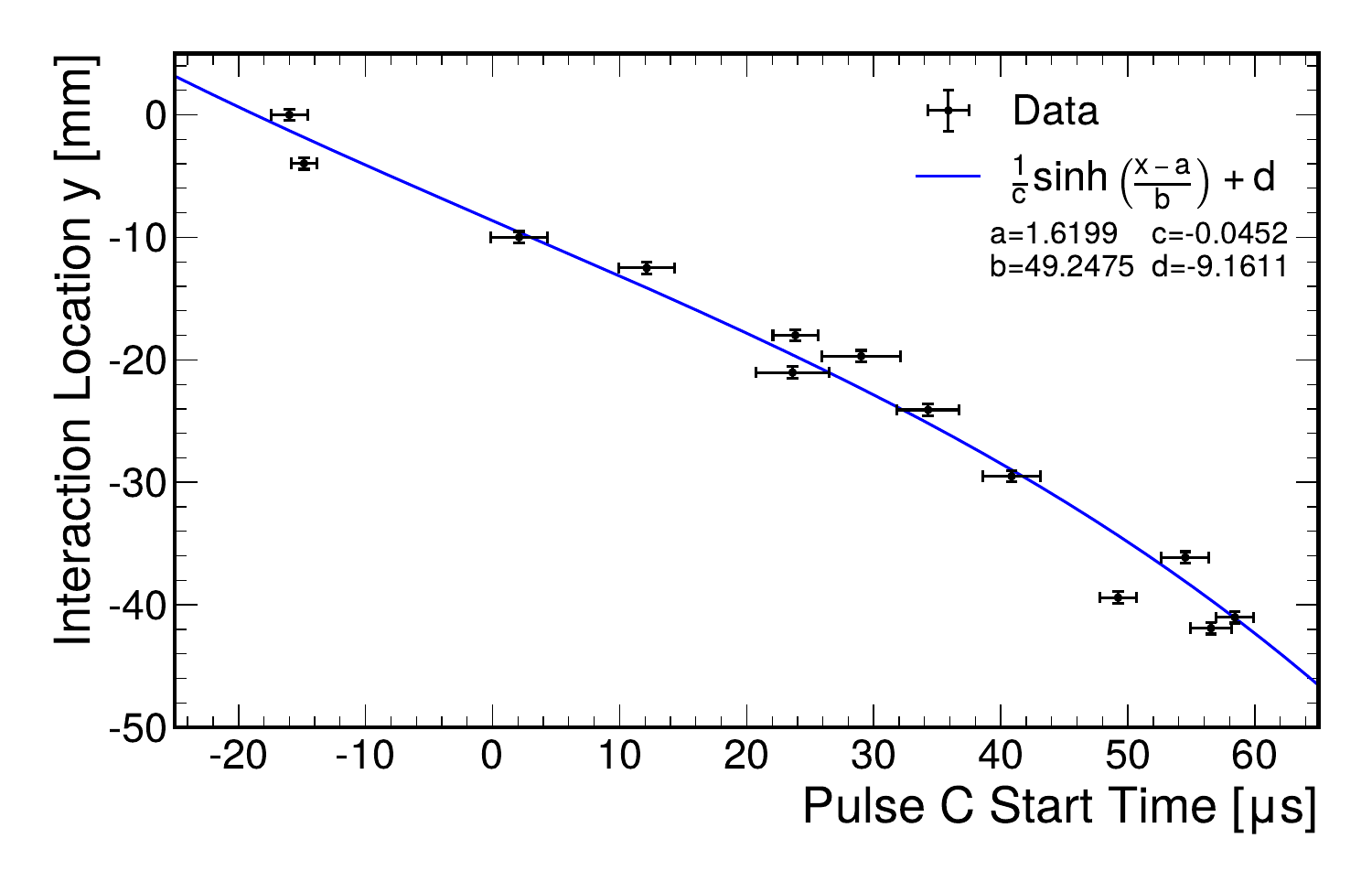}
    \caption{Interaction location as a function of a single pulse-timing parameter. Pulse C Start Time is defined as the difference in time between when Pulse C rises to 30\% of its peak and when Pulse A (used as a reference) rises to 30\% of its peak. The points are fit to a simple sinh(x) function (in blue). The horizontal error bars are the standard deviations for all interactions identified as coming from that location.
}
    \label{fig:Start}
\end{figure}

In CDMS, a variety of timing parameters have been used to determine impact position~\cite{SCDMS2018,CDMSLite}.  Once such parameters are formed, position information can be derived by comparing source position with the parameter and fitting to a function.  Such an approach can be applied to this dataset as well, which can provide an analytical benchmark with which to compare ML results.  An example of such a timing parameter is the time when the pulse amplitude in channel C reaches 30\% of its peak amplitude compared to the same point in relative amplitude of the channel A pulse. 
The data can be fit to a hyperbolic sine function in order to predict the location of an
interaction at any unknown point along this line (Figure~\ref{fig:Start}). 

Examining Figure~\ref{fig:locations} reveals why the start time of Pulse C is a good choice for a single parameter to characterize position in this dataset: the source was moved along a path taking it farther and farther from Channel C. After fitting using to the model-learning subset (see~\cref{sec:ml-data}) we obtain a root mean squared error (RMSE) of $2.125 \mathrm{mm}$ on the held-out subset points. This example curve fit provides an analytical benchmark with which to compare machine learning solutions. 


While we see some success from fitting from a single variable in channel C, there is clearly much more information in the waveforms of all the different channels.
Finding the optimal use of all the information is a task better suited to machine learning. 
The problem can be formulated as a standard regression task where the relationship between the impact location ($y$) and the collection of observed pulse information ($\bm{x}$) can be expressed as
\begin{equation}\label{eq:model_form}
    y = f_{\bm{\zeta}}(\bm{x}; \theta)
\end{equation}
where $f_{\bm{\zeta}}$ represents the surrogate ML model parameterized by trainable parameters $\theta$. The model itself can subject to a set of tunable hyperparameters, represented by ${\bm{\zeta}}$. In the following section, we describe the datasets prepared for training and evaluating ML models to perform the regression task described in~\cref{eq:model_form}.

\section{Machine Learning Techniques}

\subsection{Classical Regression Techniques}
In order to establish a benchmark of performance metrics and feature importance, linear regression is used as our first attempt to solve the position reconstruction problem.  Being one of the simplest regression techniques, linear regression often provides a quick and robust estimate for input-output relationships for small-to-medium sized datasets. 
In this model, the reconstructed position $(y)$  is expressed as a linear combination of the input features $(\vec{x})$ 
\begin{equation}
y = \theta_0 + \sum_{i=1}^k \theta_i x_i
\label{eqn:linreg}
\end{equation}
where $k$ is the number of input features.  The parameters of the model (the bias term $\theta_0$ and the $\theta_i$ coefficients) are determined by minimizing the mean squared error (MSE), often accompanied by a regularizing term
\begin{equation}
    \ell = \frac{1}{N}\sum_{j=1}^N \left(y_j - \hat{y_j} \right)^2 + \alpha R_p\left(\vec{\theta}\right)
    \label{eqn:linregloss}
\end{equation}
where $N$ is the number of training samples, $y$ and $\hat{y}$ respectively represent the true and predicted values of the location,   $R_p\left(\vec{\theta}\right) = \sum_i |\theta|^p$ is the regularization term, and $\alpha$ is the regularization strength. The ordinary least squared (OLS) method sets $\alpha = 0$ while the Ridge and Lasso regularizations set $p=2$ and $p=1$ respectively with $\alpha > 0$.  $\mathrm{RMSE} = \sqrt{\frac{1}{N}\sum_{j=1}^N \left(y_j - \hat{y_j} \right)^2} $ is considered as the performance metric. 

\begin{figure*}[!htpb]
\centering
\subfigure[]{ {\includegraphics[width=0.48\textwidth]{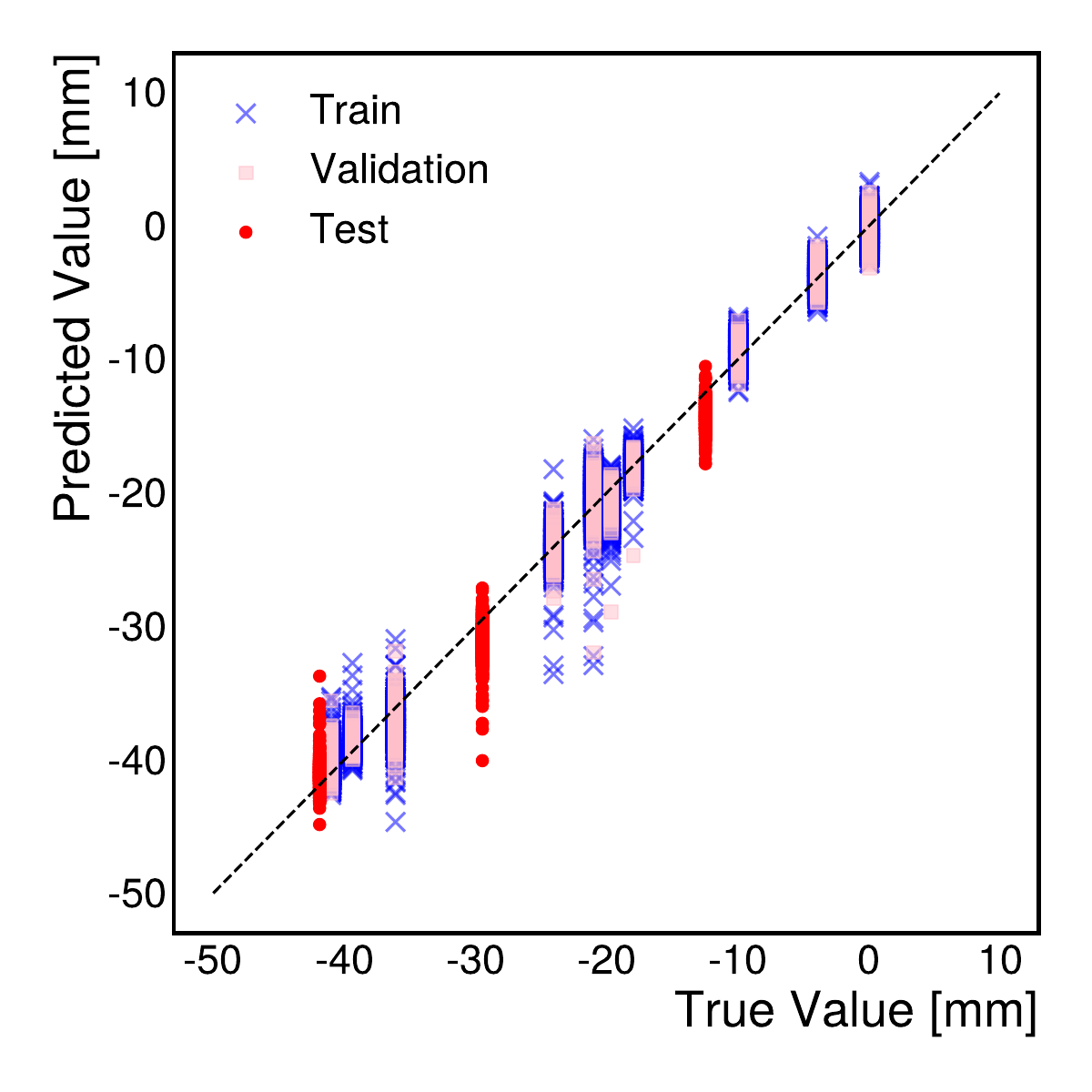}} 
\label{fig:S_Reco}}
\subfigure[]{ {\includegraphics[width=0.48\textwidth]{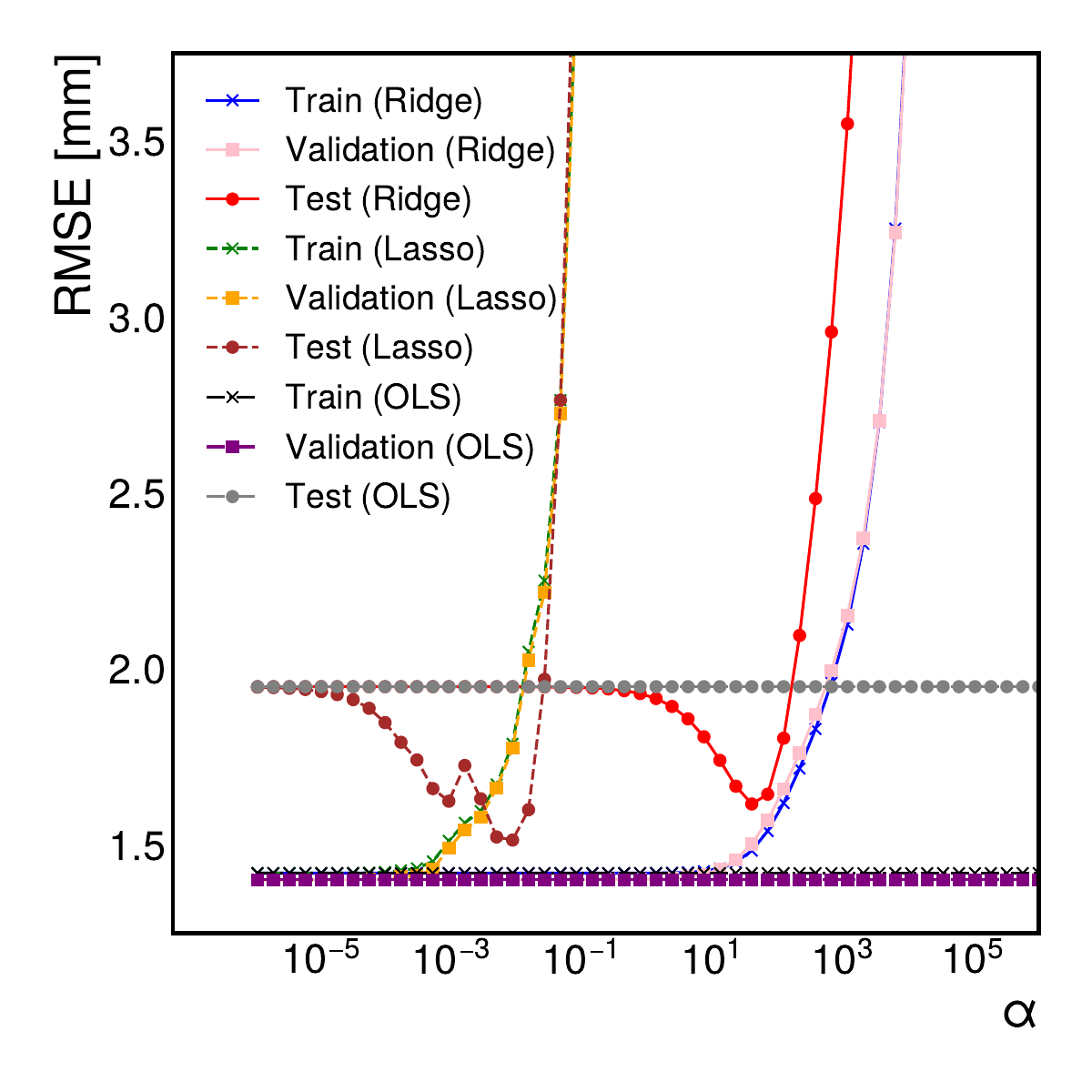}} 
\label{fig:S_Scan}}
\caption{Comparison of the predicted and true interaction positions from the OLS model (left) and comparison of the MSE losses from the training, validation, and test data from OLS regression with those from the Ridge and Lasso regression models for different values of the regularization parameter $\alpha$ (right).}
\label{fig:S_Reco_Scan}
\end{figure*}

\begin{figure*}[!htpb]
\centering
\subfigure[]{ {\includegraphics[width=0.48\textwidth]{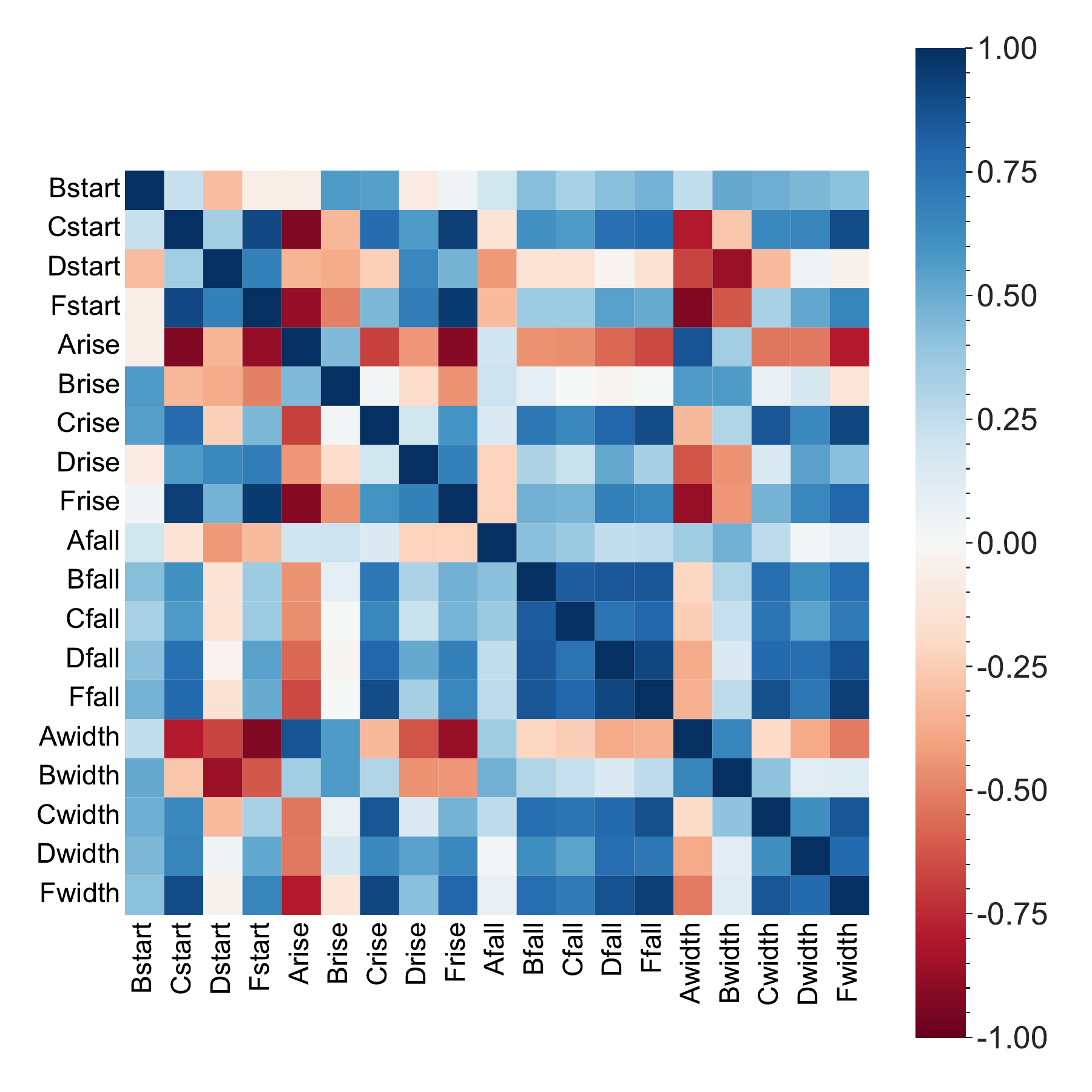}} 
\label{fig:corr-s-train}}
\subfigure[]{ {\includegraphics[width=0.48\textwidth]{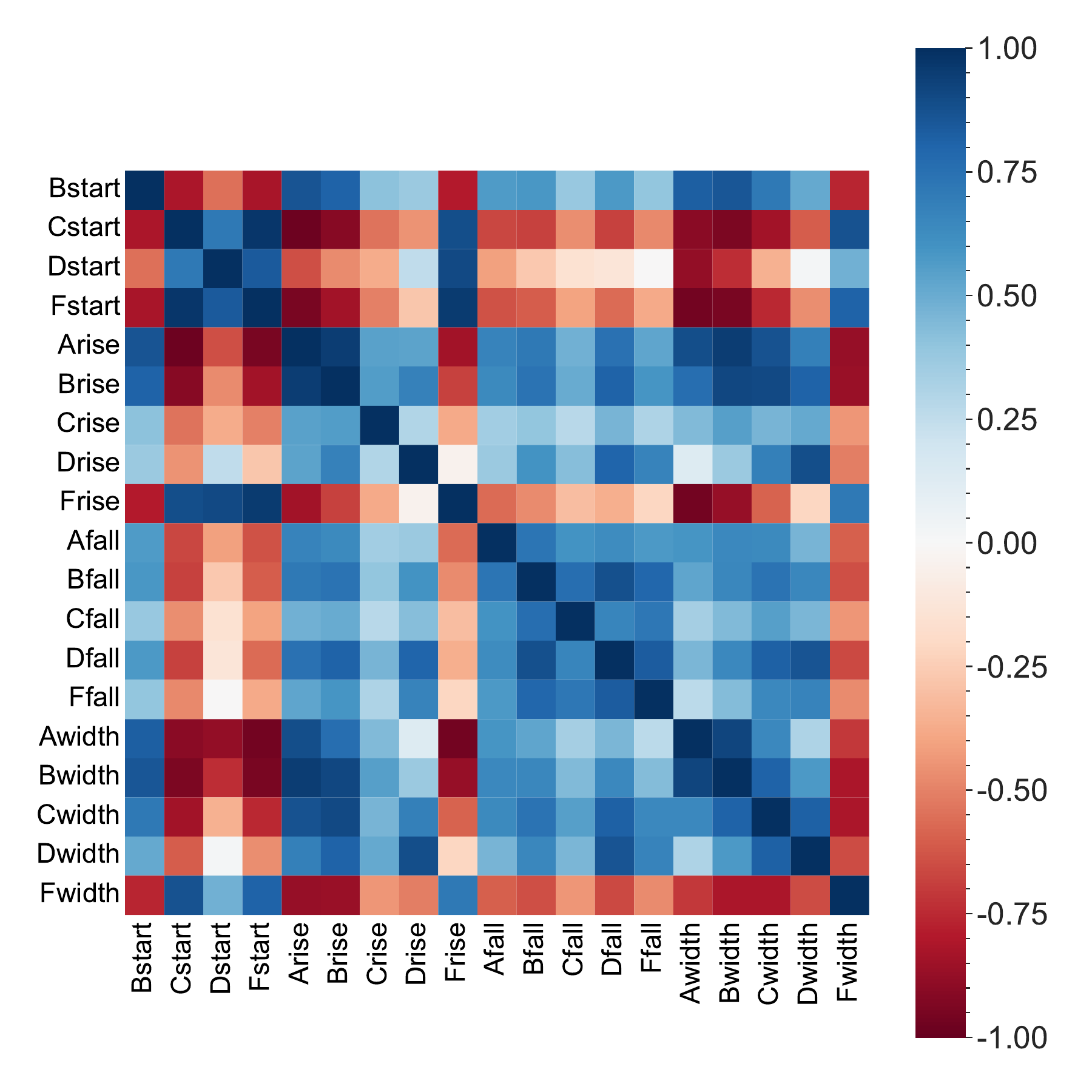}} 
\label{fig:corr-s-test}}
\caption{Correlation among input features of the simple dataset for \protect\subref{fig:corr-s-train} the training dataset and \protect\subref{fig:corr-s-train} the test dataset.}
\label{fig:S_Corr}
\end{figure*}

Figure~\ref{fig:S_Reco} shows the comparison of true and predicted values of the interaction location for the OLS regression on the reduced dataset. In figure~\ref{fig:S_Scan}, the RMSE values for the training, validation, and test data are compared with those from Ridge and Lasso regression models for different choices of $\alpha$ for the same dataset. One immediate observation is the model's poor performance in generalizing for the test data: the RMSE  from OLS for the test data is much larger than those for the training and validation subsets. Also, the values of $\alpha$ that have the least RMSE values for the test data in Ridge and Lasso regressions don't have the smallest RMSE values for training and validation data. This suggests that the model does not generalize well for unobserved impact locations. The model's inability to generalize well can be attributed to the differences in the feature correlations between the training and test data. As shown in~\cref{fig:S_Corr}, the correlation structure is notably different between training and test data, hinting at the possibility that feature correlation itself may be a function of the impact location.

Since the simple dataset shows a visibly distinct correlation structure among its features, thus prohibiting the model to generalize, it is instructive to investigate if this problem can be overcome using the extended dataset.
We found that the same training and test subsets showed a more consistent correlation structure for the timing information in the extended dataset.
Therefore, the performance gap between the training and test set might become smaller by training a regression model on the extended dataset. 
In order to identify the combination of dataset and model that gives the best performance and generalization, the behavior of the OLS, Ridge, and Lasso regression models was examined for (a) the reduced dataset, (b) the extended dataset without the amplitude information, and (c) the extended dataset with the amplitude information. Additionally, owing to the large correlations among different feature pairs in these datasets, principal component analysis (PCA) was applied to reduce the number of features and remove multi-colinearity from the input. A subset of the principal components was chosen, which accounted for 99.9\% of the observed variance in the training data.  These reduced features were used in place of the original features, $\bm{x}$, in the linear regression model in Eqn.~\ref{eqn:linreg}.

\begin{figure*}[!htbp]
\begin{center}
{\includegraphics[width=0.85\textwidth]{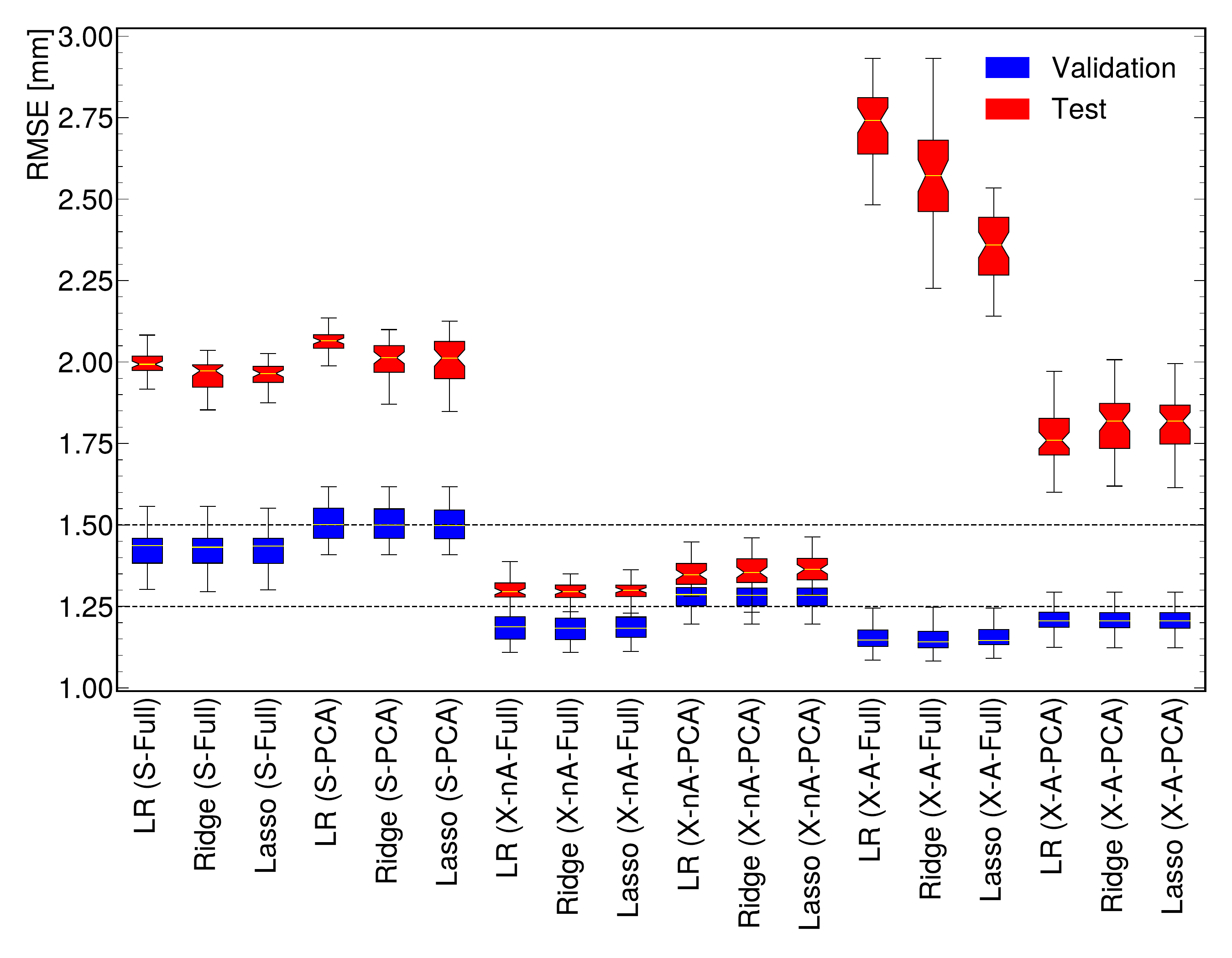}} 
\end{center}
\caption{The distribution of RMSE values for validation and test data for different choices of dataset and fit models. The different labels along the X axis refer to the different data-model combinations. 
\texttt{S, X-nA,} and \texttt{X-A} refer to the reduced, extended without amplitude, and extended with amplitude datasets respectively.
\texttt{Full} and \texttt{PCA} refer to the models where the full feature-set and the PCA-reduced features are used respectively.
The distributions of the RMSE losses were obtained for 50 different choices of random splitting for training and validation dataset.}
\label{fig:losssummary}
\end{figure*}

\begin{figure*}[!htbp]
\begin{center}
{\includegraphics[width=\textwidth]{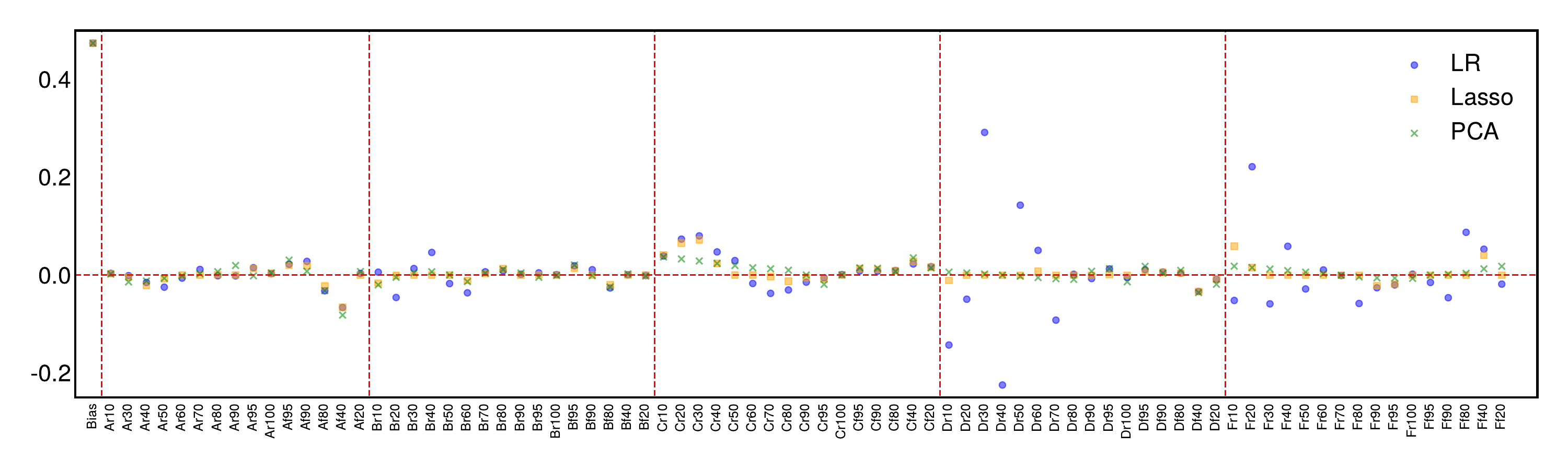}} 
\end{center}
\caption{The coefficients associated with different features of the extended dataset without pulse amplitude from OLS, Lasso, and PCA-transformed regression models. Features enclosed within two successive vertical red lines relate to the same detector channel. The Lasso regression was performed with $\alpha = 10^{-4}$.
}
\label{fig:coefcomp}
\end{figure*}

Figure~\ref{fig:losssummary} shows the distribution of the RMSE values for different choices of dataset and associated models. Besides providing a comprehensive overview of model performance for different datasets and regression models, it also allows us to make some important observations. First, including the amplitude information significantly impairs the model's ability to generalize to the test data. The known issues with calibration and amplitude measurement as described in the previous section may be responsible for this behavior. Second, the best performance is obtained from the extended model without the amplitude information. Not only the validation and the test RMSE are much lower than the ones obtained from the reduced datasets, the test-data RMSE values are comparable with the validation-data RMSE which indicates the model's ability to generalize well. Third, the spread of RMSE values for this dataset is also relatively small, indicating its insensitivity to the actual training-validation split. And finally, the performance of the OLS, Ridge, and Lasso models are comparable for both full and PCA-transformed feature sets, but the PCA-transformed features always show worse performance than the full dataset. 

While the performance metrics suggest that the OLS regression on the extended dataset without amplitude information performs as well as the regularized variants, the actual estimate of coefficients can be unstable since there are a large number of highly correlated feature-pairs in the dataset (Figure~\ref{fig:coefcomp}). The Lasso regularization helps reduce the number of effective features in the model since it allows setting the coefficients of \textit{unimportant} features (e.g. features that don't offer independent information) to zero. Examining the coefficients of the Lasso regression with $\alpha = 10^{-4}$, we found that between 33--37 of the 79 coefficients were set to zero for different training/validation splits. The number and the set of coefficients actually set to zero varied between iterations. This is understandable since the model can obtain the same information from different linear combinations of highly-correlated features. Some of the features with coefficients of larger magnitude from the Lasso regression received weaker coefficient estimates from the PCA-transformed regression, which accounts for the  degradation of performance in PCA-transformed models. 


To identify a subset of important features in a robust fashion, we applied a second method for feature selection using a variance inflation factor (VIF). VIF is a feature-wise metric that determines how well as feature can be expressed as a linear combination of other features, defined as
\begin{equation}
    \mathrm{VIF}_j = \frac{1}{1 - R_j^2}
    \label{eqn:VIF}
\end{equation}
where $R_j^2$ is the goodness-of-fit measure (i.e. the $R^2$ value) for the $j$-th feature. Large values of VIF indicate that a feature is highly correlated with other features in the dataset.  To obtain a reduced set of features for the extended dataset without amplitude, we sequentially removed features from the dataset which showed VIF~$> 1000$. This sequential feature pruning allowed simplification of the dataset while keeping a subset of  relatively independent features. The OLS model was then fit using the reduced feature-set. A $t$-test was performed to determine if the coefficient estimates were significant at $95\%$ confidence level. To account for the variability due to training/validation split, this procedure was performed $N_E = 50$ times with different random splittings of the training/validation data, thus providing an estimate of the importance of the features using the following relation:
\begin{equation}
    \mathrm{Importance}[j] = \frac{1}{N_E}\sum_{i=1}^{N_E} |\theta_j|_i \cdot [\mathrm{VIF}_j < 1000]_i \cdot [p_j < 0.05]_i
    \label{eqn:imp}
\end{equation}
where the index $i$ represents the different iterations of the experiment and $p_j$ is the $p$-value associated with the $t$-test on the $j$-th feature. It should be noted that the importance assigned to a feature using Eqn.~\ref{eqn:imp} depends the actual ordering of the features on the dataset. While the process determines \textit{one} subset of features that can describe the data well, a different subset can be chosen based on a different ordering of the input features.

\begin{figure*}[!htbp]
\begin{center}
{\includegraphics[width=0.5\textwidth]{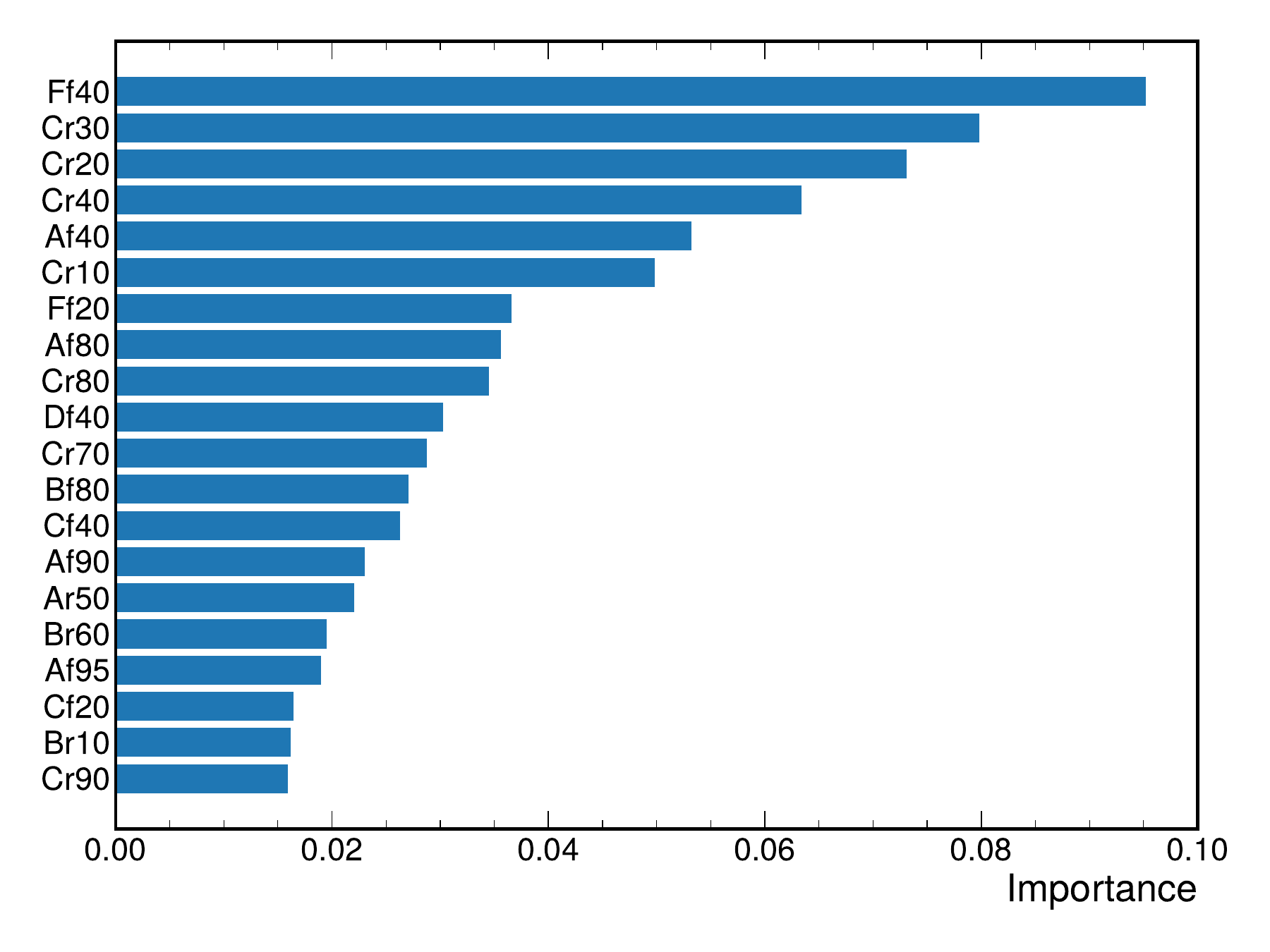}} 
\end{center}
\caption{The top 20 features and the corresponding importance values (~\cref{eqn:imp}) for the extended dataset without amplitude information.
}
\label{fig:impsummary}
\end{figure*}

Figure~\ref{fig:impsummary} shows the top 20 features and the importance values assigned to them. Note that the reduced model's performance gets worse for VIF thresholds less than 1000. Also, the number of features that were assigned zero importance by Eqn.~\ref{eqn:imp} was 35, similar to the number of zero coefficients found from the Lasso regression model. Comparing~\cref{fig:coefcomp} with~\cref{fig:impsummary} reveals that the most important features following~\cref{eqn:imp} also have relatively larger coefficients from the Lasso regression. The median validation and test RMSE for this reduced model were 1.24 and 1.32 respectively, close to the numbers obtained from the OLS and Lasso regression models.


\subsection{Deep Neural Networks}

 Deep neural networks (DNN), which operate using a sequence of matrix operations, can be understood as a natural extension of the linear processes from the previous classical regression approach. The key advantage of DNN is the presence of non-linear activation functions and an increased number of layers and trainable parameters. These attributes enable neural networks 
to identify more complex relationships between input and output data that may be inaccessible through linear regression.  Using neural networks, the process of learning the mapping function $f$ from~\cref{eq:model_form} is transformed into  solving the optimization problem
\begin{equation}
   \min_{\theta} \; \ell(y, f(\bm{x};\theta)),
\end{equation}
where $\bm{x}$ is the extracted features from observed signals, $y$ is the true position/location (ground truth), $\ell$ is the loss function, and $\theta$ is the network weights. For our implementations, we used root mean squared error (RMSE) as our loss function, defined as the square root of the MSE from~\cref{eqn:linregloss}.

\subsubsection{Preliminary work on DNN models}


As an initial exploration of deep neural networks for this task, three fully-connected multi-layer perceptron (MLP) networks were implemented, with 2, 5, and 10 hidden layers each with 32 nodes.
Each hidden layer was followed by 
a batch normalization~\cite{ioffe2015batch} layer,
a leaky rectified linear (LeakyReLU) activation~\cite{maas2013rectifier,xu2015empirical}, and
a dropout layer~\cite{JMLR:v15:srivastava14a}.  The dropout layer was assigned a dropout rate of 0.5 to prevent over-fitting the network due to highly correlated input features. 
For the output layer, the learned features were simply passed through a linear layer and its prediction was obtained directly.



The Adam~\cite{kingma2014adam} optimizer was used, with a constant learning rate of $0.001$. 
Each network was trained for up to $500$ epochs.
Since network performance varies due to the random initialization of the network weights, 50 iterations of our network were trained with randomly chosen seeds to observe the average performance of our network architecture and training method. 

\begin{figure*}[!htpb]
\centering
\subfigure[DNN with 2 hidden layers]{ {\includegraphics[width=0.3\textwidth]{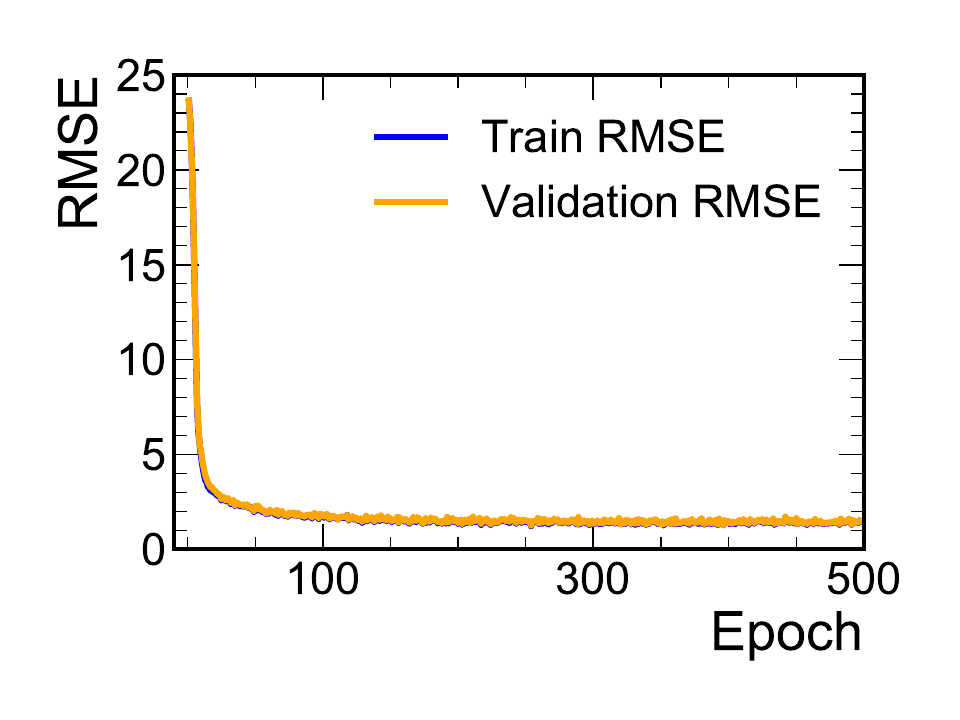}} }
\subfigure[DNN with 5 hidden layers]{ {\includegraphics[width=0.3\textwidth]{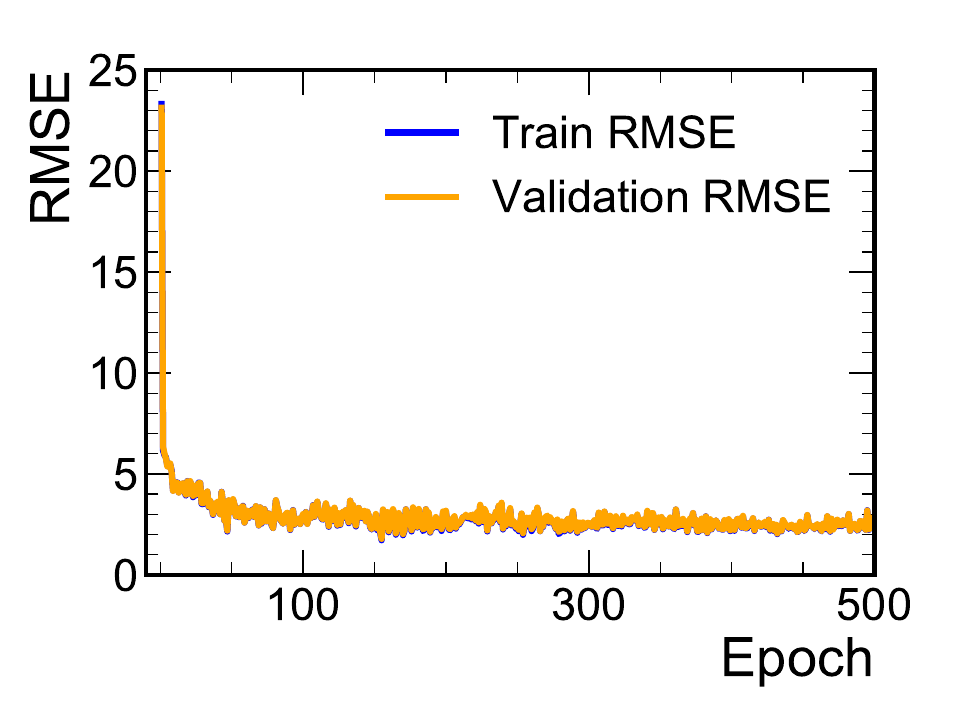}} }
\subfigure[DNN with 10 hidden layers]{ {\includegraphics[width=0.3\textwidth]{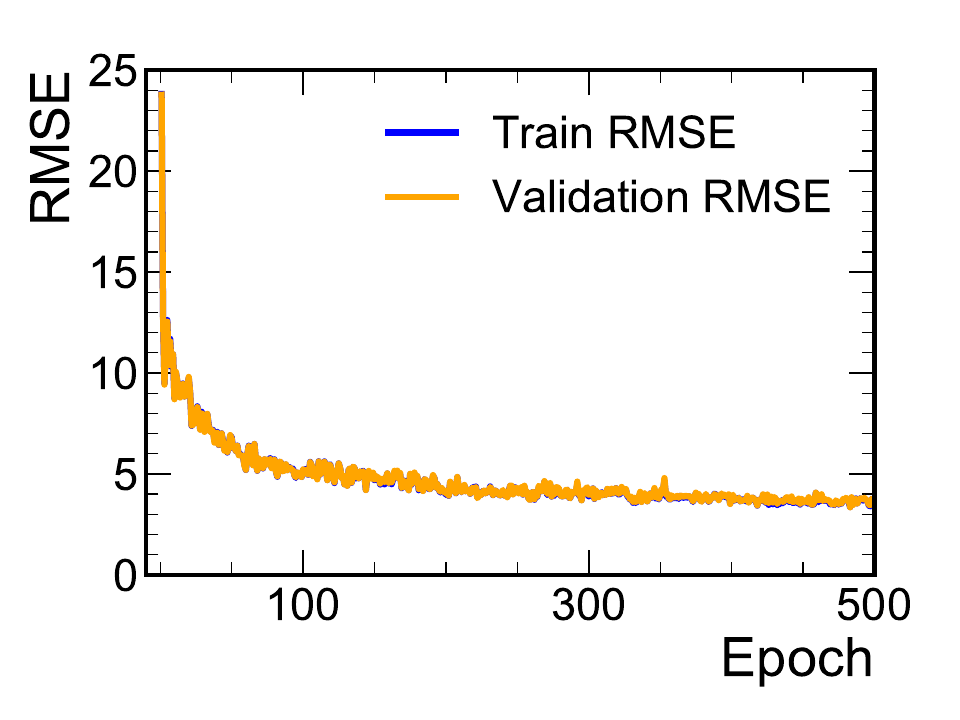}} }
\caption{The performance (RMSE) curves on training and validation set.
}
\label{fig:traing_curves}
\end{figure*}

The training and validation performance curves are shown in~\cref{fig:traing_curves} for models with 2, 5, and 10 hidden layers. Note that all three settings converged well before $500$ epochs and do not show over-fitting issues between the training and validation data. 
As expected, the neural network model with 10 hidden layers takes the longest training time to converge due to increased model complexity.

\begin{figure*}[!htbp]
\begin{center}
{\includegraphics[width=0.75\textwidth]{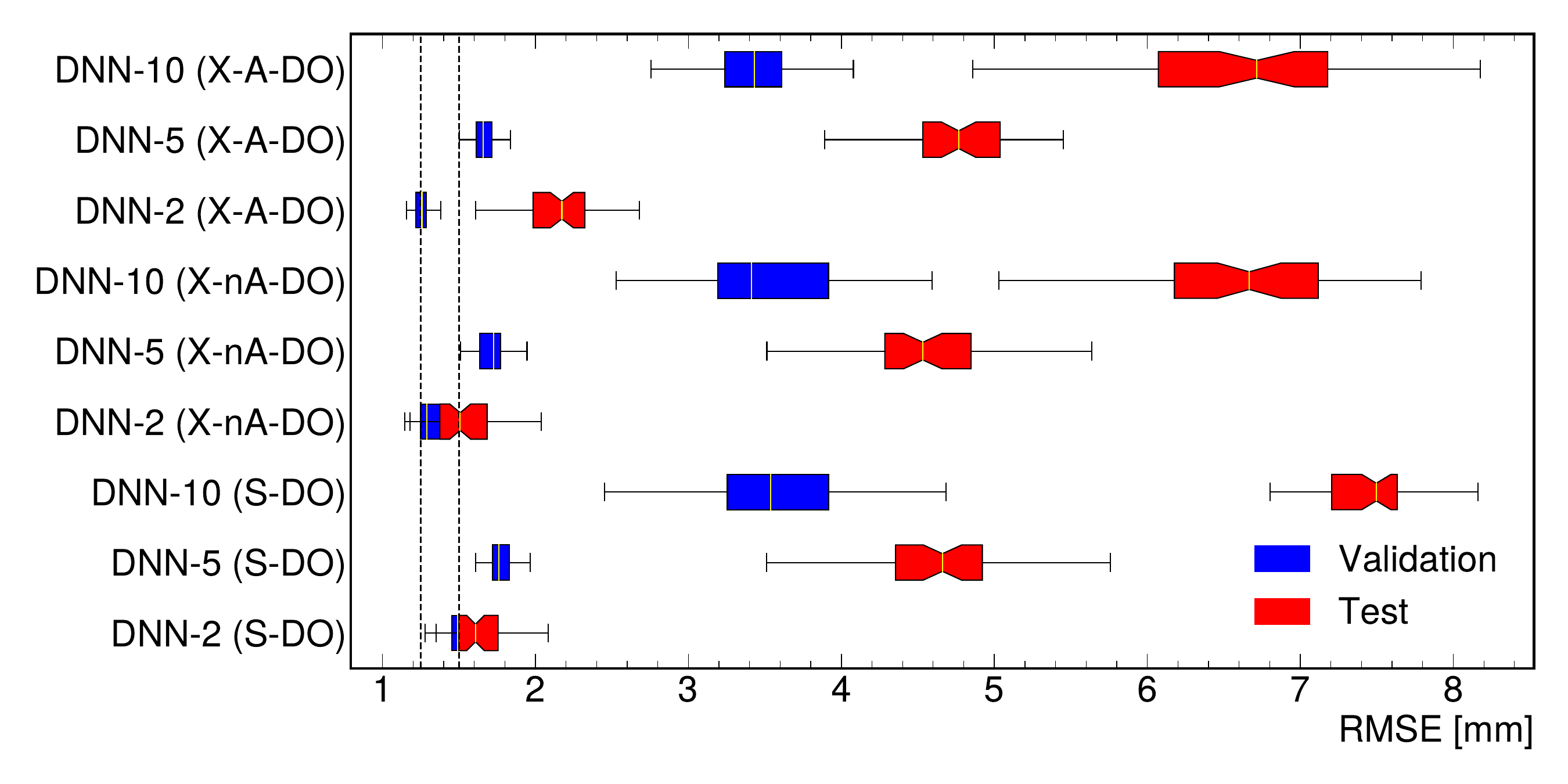} 
}
\end{center}
\caption{The distribution of loss values for validation and test data for different choices of data set and DNN models using dropout regularization. The different labels along the X axis refer to the different data-model combinations. 
\texttt{S, X-nA,} and \texttt{X-A} refer to the reduced, extended without amplitude, and extended with amplitude data sets respectively.
\texttt{DNN-2, DNN-5,} and \texttt{DNN-10} refer to DNN models with 2, 5, and 10 hidden layers respectively. 
The distributions of the RMSE losses were obtained for 50 different choices of random initialization of network weights.}
\label{fig:DNNLossSummary}
\end{figure*}

The performance of our neural network models on the three datasets: (1) \textit{simplified}, (2) \textit{extended without amplitude}, and (3) \textit{extended with amplitude} is summarized in~\cref{fig:DNNLossSummary}. 
Similar to what was observed in case of the linear regression models, DNN models trained on the extended dataset without amplitude information give us the best performance in terms of its ability to close the performance gap between the training and test sets.
Because of the calibration issues with the recorded pulse amplitude information in the full data set as noted in~\cref{sec:scan}, inclusion of this information significantly worsens the models' generalizability.

\begin{figure*}[!htpb]
\centering
\subfigure[]{
\includegraphics[width=0.45\textwidth]{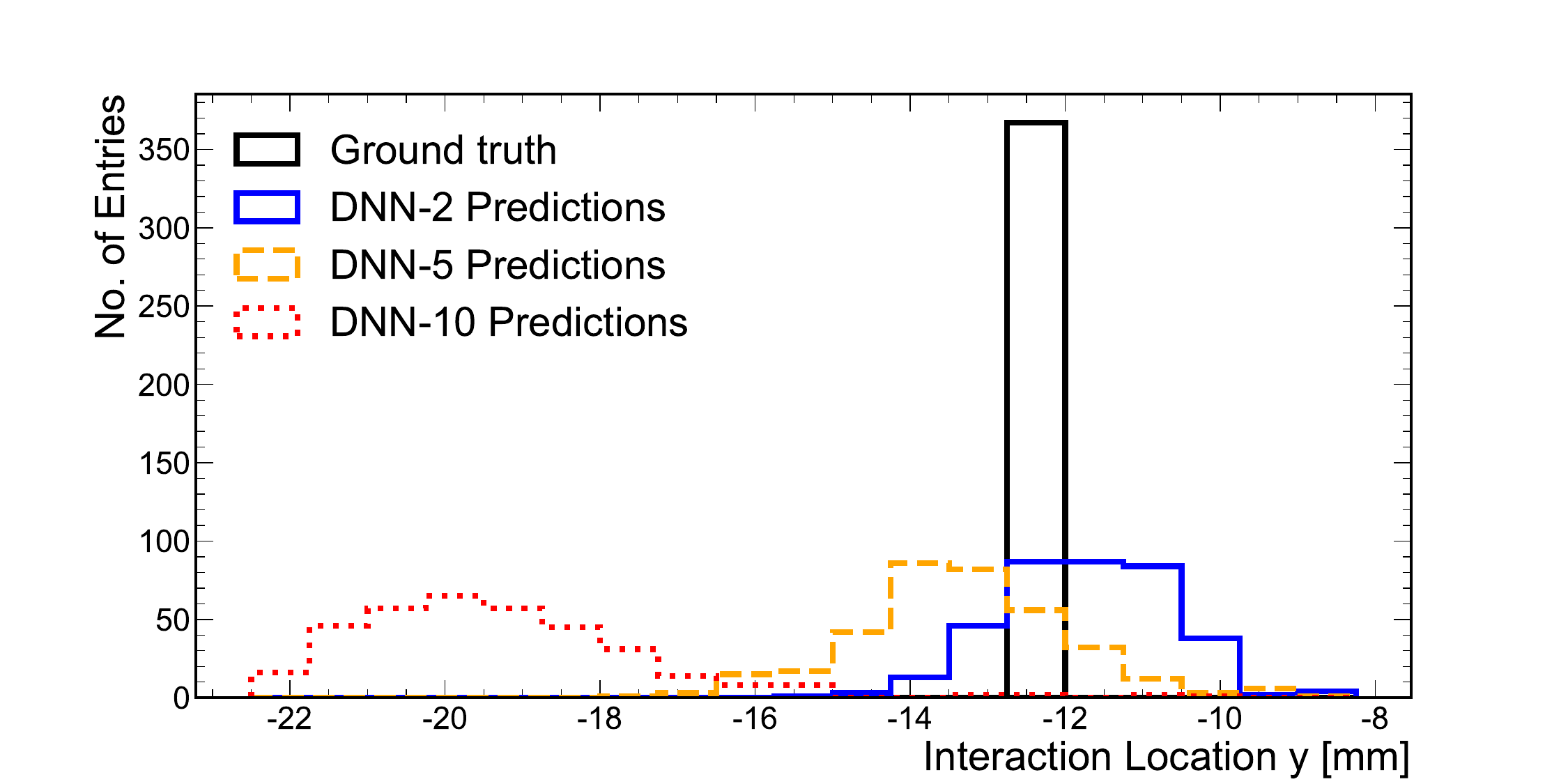}
\label{fig:hist_12.502}
}
\subfigure[]{
\includegraphics[width=0.45\textwidth]{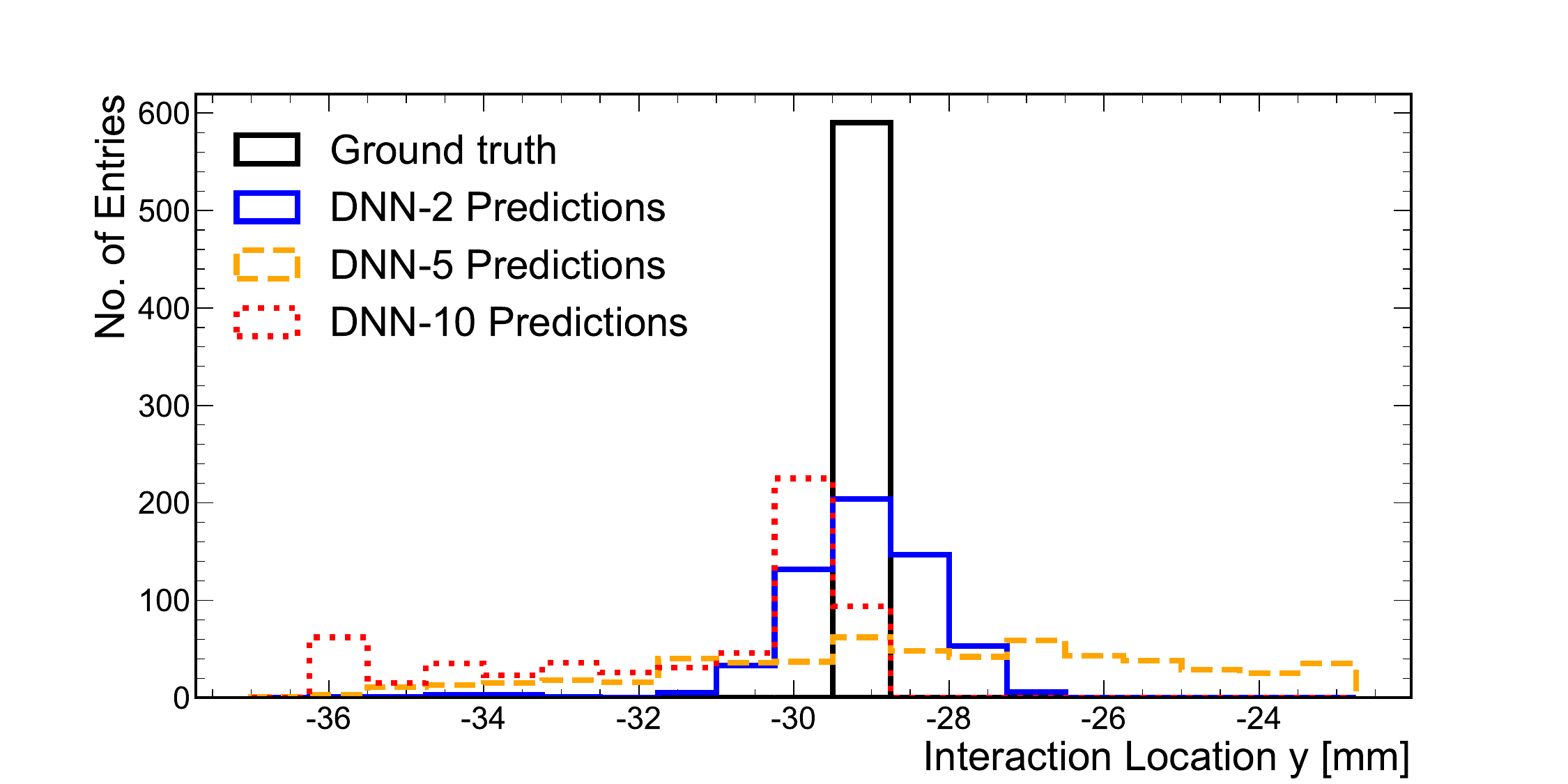}
\label{fig:hist_29.500}
} \\
\subfigure[]{
\includegraphics[width=0.48\textwidth]{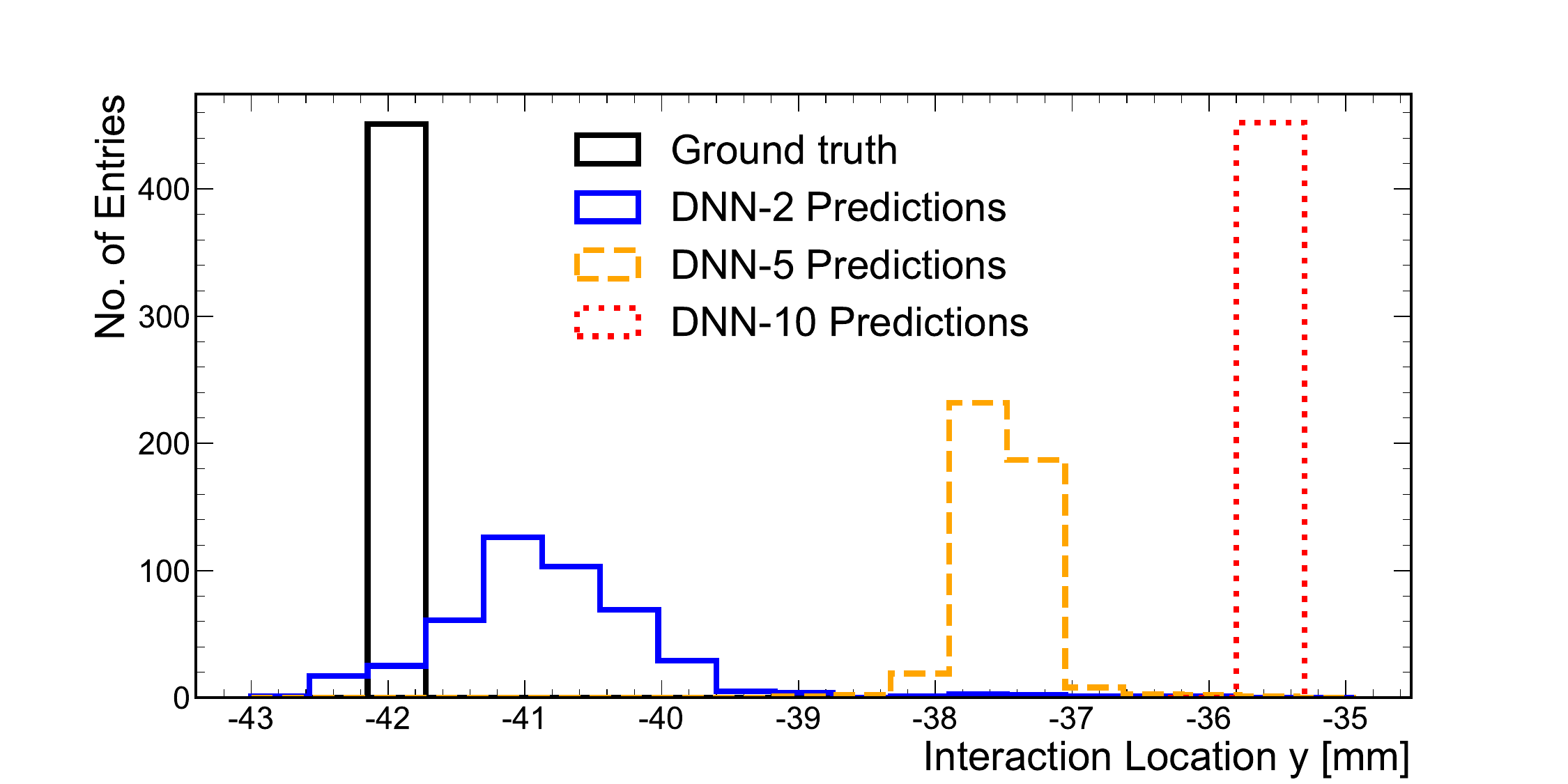}
\label{fig:hist_41.900}
}
\caption{Distribution of predicted impact location from the DNN models with 2,5, and 10 hidden layers where the true impact location is 
\protect\subref{fig:hist_12.502} --12.502~cm, 
\protect\subref{fig:hist_29.500} --29.5~cm, and
\protect\subref{fig:hist_41.900} --41.9~cm.
For each DNN architecture, the trained model with the minimum RMSE value on the HOS is chosen.}
\label{fig:hist_HOS}
\end{figure*}

 Contrary to expectations, naively increasing the model complexity does \textbf{not} improve the performance. Instead, not only do we see an increase
in the RMSE values for larger network architectures, their ability to generalize for
the Held-out test set (HOS) worsens increasingly with the number of hidden layers. 
This is also verified in the distributions of prediction from  DNN-2, DNN-5, and DNN-10 models trained on the simple dataset for the three impact locations in the HOS, as shown in~\cref{fig:hist_HOS}. Only the predictions made by the iteration of each DNN which produced the smallest RMSE for the test data is shown. It can be seen that DNN-2 yields the best predictions across all test positions, consistent with the results in~\cref{fig:DNNLossSummary}.

\subsubsection{Exploring larger neural networks}

    In the previous section, we observed that increased network complexity resulted in worse performance on the held-out set data, with the largest network DNN-10 suffering the most.  
    This result can be at least partly attributed to the use of dropout as the primary means of preventing over-fitting. The effectiveness of dropout to control overfitting highly depends on the size and quality of the dataset, as well as the choice of dropout rate as a function of the model architecture~\cite{srivastava2013improving,pauls2018determining}. Using the same drop ratio for increasing sizes of networks can be detrimental to network performance since larger networks will take a longer time and would require a much larger training set to optimize fitting given their increased complexity. As a result, the drop ratio must be carefully readjusted for larger networks to ensure that different components don’t drop out too early. 

    Following these observations, we explored whether an alternative method to combat over-fitting would be more effective for larger network architectures, possibly allowing us to achieve better performance from more complicated network architectures. Rather than readjusting the drop ratio for larger networks to find an optimal value through a process of trial and error, the dropout strategy was replaced with a new training method using L1 regularization and an early stopping procedure, referred to as \textbf{L1ES}.  As discussed in~\cref{eqn:linregloss}, L1 or Lasso regularization imposes a penalty term to the loss function of our regression function that is proportional to the sum of the absolute values of weights, and biases with regularization strength parameter $\alpha$, set to $\alpha=1$ in this study. By adding this term to our loss function, the network is encouraged to have smaller weights. This in turn results in our network being less sensitive to variations in input data and therefore less prone to overfitting. 
    In addition to L1 regularization, an early stopping procedure was implemented as follows. The training and validation loss was monitored during the training procedure. If the training loss decreased while the validation loss increased, the training automatically concluded at that epoch and the network's configuration was saved. This early stopping procedure tends to result in a more stochastic training process compared to models trained with dropout. Additionally, batch normalization and L1 regularization try to regularize the network with conflicting approaches~\cite{vanlaarhoven2017l2}, which was confirmed by our early attempts showing significantly worse network performance. Thus, batch normalization was removed from the network architecture for all models trained with L1 regularization.
    
    To explore the performance of larger networks with this new training method, the DNN-10 model from the previous section was trained using the L1ES method. Additionally, two new network architectures were introduced. 
    The first new network model is a Large Dense Neural Network (LargeDNN) that acts as an upscaled MLP with a variable number of nodes for the hidden layers. The LargeDNN model consisted of 5 hidden layers with 100, 100, 50, 25, and 10 nodes in them respectively.
    Our second novel architecture is a more complicated Convolutional Neural Network (CNN), which features a one-dimensional convolutional input layer.

   \begin{table}[!htpb]
    \caption{Summary of network architectures and the number of trainable parameters for each of the NN-based models used in this study. Parameter counts are computed assuming the reduced data set as the input.}
    \begin{center}
    \begin{tabular}{llr}
    \toprule
         Model Name & Network Architecture \hspace{30pt}& \hspace{30pt} Parameter No. \\
         \midrule
         DNN-2 & Dense & 737\\
         DNN-5 & Dense & 4,097\\
         DNN-10 & Dense & 9,697\\
         LargeDNN & Dense & 18,696\\
         CNN & Conv1D+GRU & 68,996\\
         \bottomrule
    \end{tabular}
    \end{center}
    \label{tab:arch_summary}
    \end{table}
    
    The output from this input convolutional layer was then passed onto a Gated Recurrent Unit (GRU) layer~\cite{chung2014empirical}, which is often paired with CNNs for higher-complexity data prediction tasks, before passing through a series of dense layers and returning our output.
   The parameter count of the LargeDNN and CNN models is shown in in~\cref{tab:arch_summary}  for comparison with the models used in the previous section.
    
\begin{figure*}[!htbp]
    \begin{center}
    {\includegraphics[width=.8\textwidth]{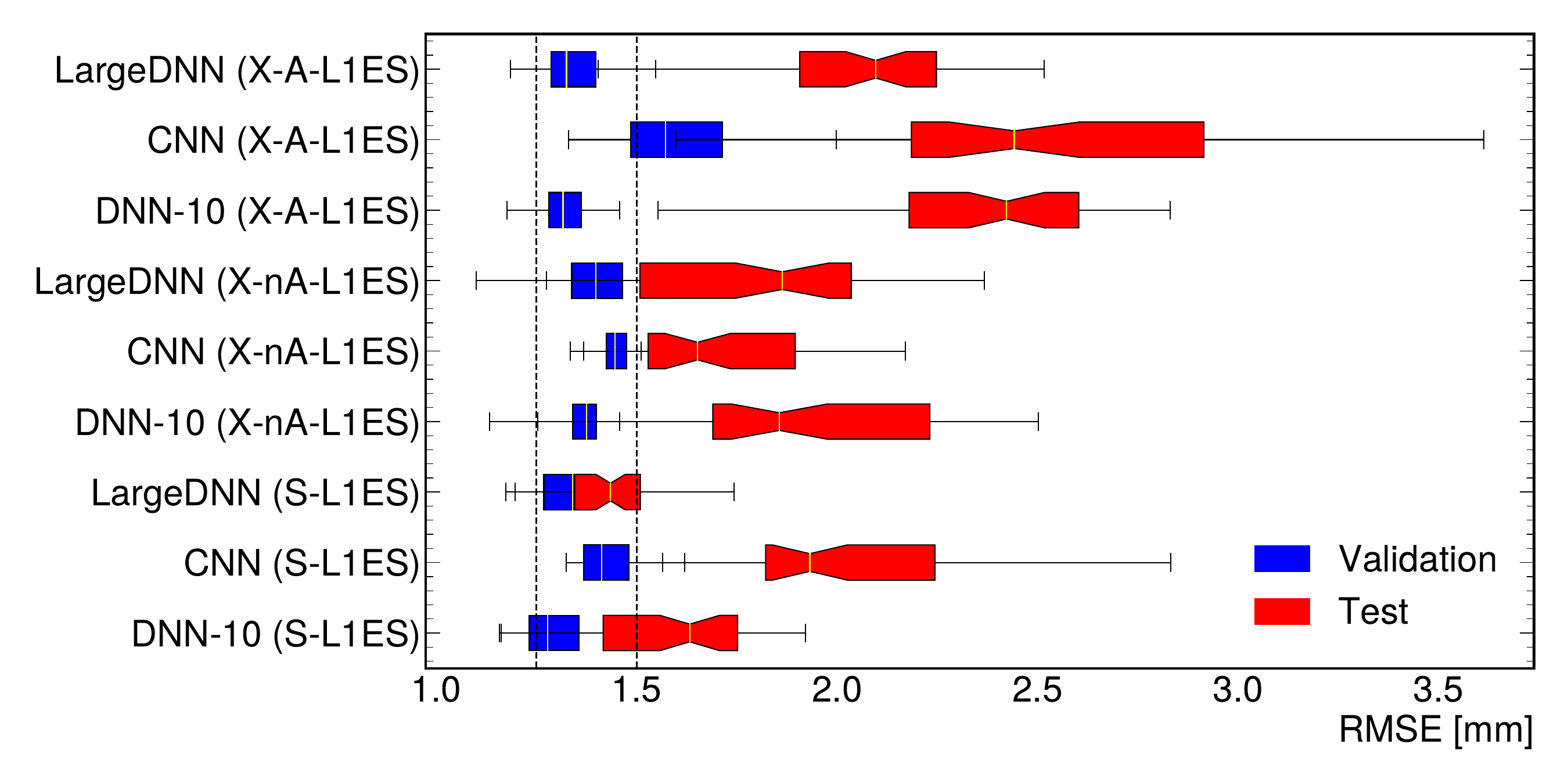} 
    }
    \end{center}
    \caption{The distribution of loss values for validation and test data for different choices of data set and DNN models using L1ES regularization. The different labels along the Y axis refer to the different data-model combinations. 
    \texttt{S, X-nA,} and \texttt{X-A} refer to the reduced, extended without amplitude, and extended with amplitude data sets respectively. The distributions of the RMSE losses were obtained for 50 different choices of random initialization of network weights.}
    \label{fig:DNNLossSummary2}
    \end{figure*}

    With a new training procedure and architecture defined,  model performance on the reduced data set can be compared with our results from the previous section. As previously mentioned, all models are trained with 50 different randomly generated initialization seeds to display the average performance of our network. The distribution of RMSE values on the validation and test data for selected models trained on all three variants of the dataset is shown in ~\cref{fig:DNNLossSummary2}. 

    From these results, it is immediately clear that the L1ES procedure is more successful than dropout in closing the performance gap across training and test data for larger networks; when trained with L1ES, the DNN-10 model outperforms the DNN-2 model with dropout when trained on the simple dataset despite the DNN-10 model being significantly larger. 
    Furthermore, the LargeDNN and the CNN models also show significant improvement in performance compared to the DNN-5 and DNN-10 models in the previous section. 
    The LargeDNN model trained with L1ES provides the best performance both in terms of the  RMSE values of validation and training set as well as closing the performance gap between these two subsets.
    Finally, including the amplitude information worsens the generalization performance, similar to what has been observed with the models discussed in the previous sections.



    
    It should be noted that the CNN's performance lags behind that of other models. Due to the complexity of the network, this may be attributed to a variety of possible factors such as non-optimal kernel and stride sizes for the convolutional layer, needing to change our L1 regularization penalty value $\alpha$ when going from dense to convolutional/GRU layers, or the early stopping criteria being too strict for this larger network. This last factor can also explain the larger variation in the CNN RMSE compared to other models. Ending training prematurely with early stopping procedures is more likely to happen with larger models like the CNN, which would occasionally conclude training after only 2-3 epochs resulting in unfitted performance which skews the RMSE mean and standard deviation.

\subsubsection{Symbolic regression studies}

Following our neural network approaches, we introduce symbolic regression (SR) as an alternative machine learning method. Symbolic regression takes in a kernel of algebraic functions and our inputs and searches the resulting phase space to find the optimal combination to fit our data set using an evolutionary training method. The resulting function serves to alleviate some of the concerns with our other machine learning methods; it introduces non-linearity which is absent from our classical linear regression techniques while also avoiding some of the black-box nature associated with artificial neural networks by providing an analytic functional form.

In this section we explore the performance of algorithms generated using symbolic regression on our data using the same data splitting and evaluation metric as before. We implement our SR algorithm using PySR 0.11.11 \cite{cranmer2023interpretable}, providing the kernel of operators to our SR model given in~\cref{tab:sr_operators}. Given the limitations of our SR library for many variables, we have restricted ourselves to using only input parameters from the reduced data set. We first select our inputs from those with the highest correlation coefficient with the interaction location.

In~\cref{tab:sr_performance1} we list the performance of the resulting SR algorithms on our validation and test data, where we see that a three parameter function produced with symbolic regression has similar performance to our previous dense neural networks with several thousand trainable parameters. The resulting function,
\begin{equation}
  f(\vec{x})=-8.585\cdot \text{C}_{\text{start}} + \exp(\text{A}_{\text{rise}})-4.874\cdot \text{F}_{\text{start}}+\sin\left[\frac{1}{\sin(\text{C}_{\text{start}})-0.662} \right],  
  \label{eq:sr_three_param}
\end{equation}
seems unlikely to match a physical functional form pertaining to our system. Despite this, the performance of our SR function when compared to neural networks speaks to its use for algorithm compression.

\begin{table}[!htpb]
    \caption{Kernel of functions provided to our symbolic regression model.}
    \begin{center}
    \begin{tabular}{ccc}
         \toprule
         Binary Operators & \multicolumn{2}{c}{Unary Operators}  \\
         \hline Addition & Cosine & Inversion\\
         Multiplication & Sine & Exponentiation\\
         \bottomrule
    \end{tabular}
    \end{center}
    \label{tab:sr_operators}
\end{table}
\begin{table}[!htpb]
    \caption{Validation and test/HOS set Performance of symbolic regression algorithms for different sets of limited inputs from the reduced data set.}
    \begin{center}
    \begin{tabular}{clcc}
    \toprule
     No. of Inputs & List of Input Variables & Val. RMSE  & HOS RMSE \\
      \hline 1 & $\text{C}_{\text{start}}$ & 2.230& 2.148\\
      2 & $\text{C}_{\text{start}}$,$\text{A}_{\text{rise}}$ & 1.891& 1.768\\
      3 & $\text{C}_{\text{start}}$, $\text{A}_{\text{rise}}$, $\text{F}_{\text{start}}$& 2.060 & 1.393\\
      4 & $\text{B}_{\text{start}}, \;\text{C}_{\text{start}}, \;\text{D}_{\text{start}}, \;\text{F}_{\text{width}}$ & 1.767&1.393\\
      \bottomrule
    \end{tabular}
    \end{center}
    \label{tab:sr_performance1}
\end{table}

Following these results, we provide the complete reduced set of 19 input parameters to the SR library, allowing it to choose the parameters from which to construct a regression function, resulting in the output
\begin{equation}
    f(\vec{x})=(\text{C}_{\text{start}}+1.605)\cdot\left[\sin(\cos(\text{D}_{\text{start}}))-14.086\right]+\sin(\text{B}_{\text{start}}+\text{F}_{\text{width}})-0.299+\text{B}_{\text{start}},
    \label{eq:sr_four_param}
\end{equation}
which obtained a RMSE value of $1.767$ and $1.393$ on the validation and HOS sets respectively. The resulting algorithm uses four parameters with a performance is largely identical to that of the previous symbolic regression algorithm.

\begin{figure}[!htpb]
    \begin{center}
    \includegraphics[width=0.8\textwidth]{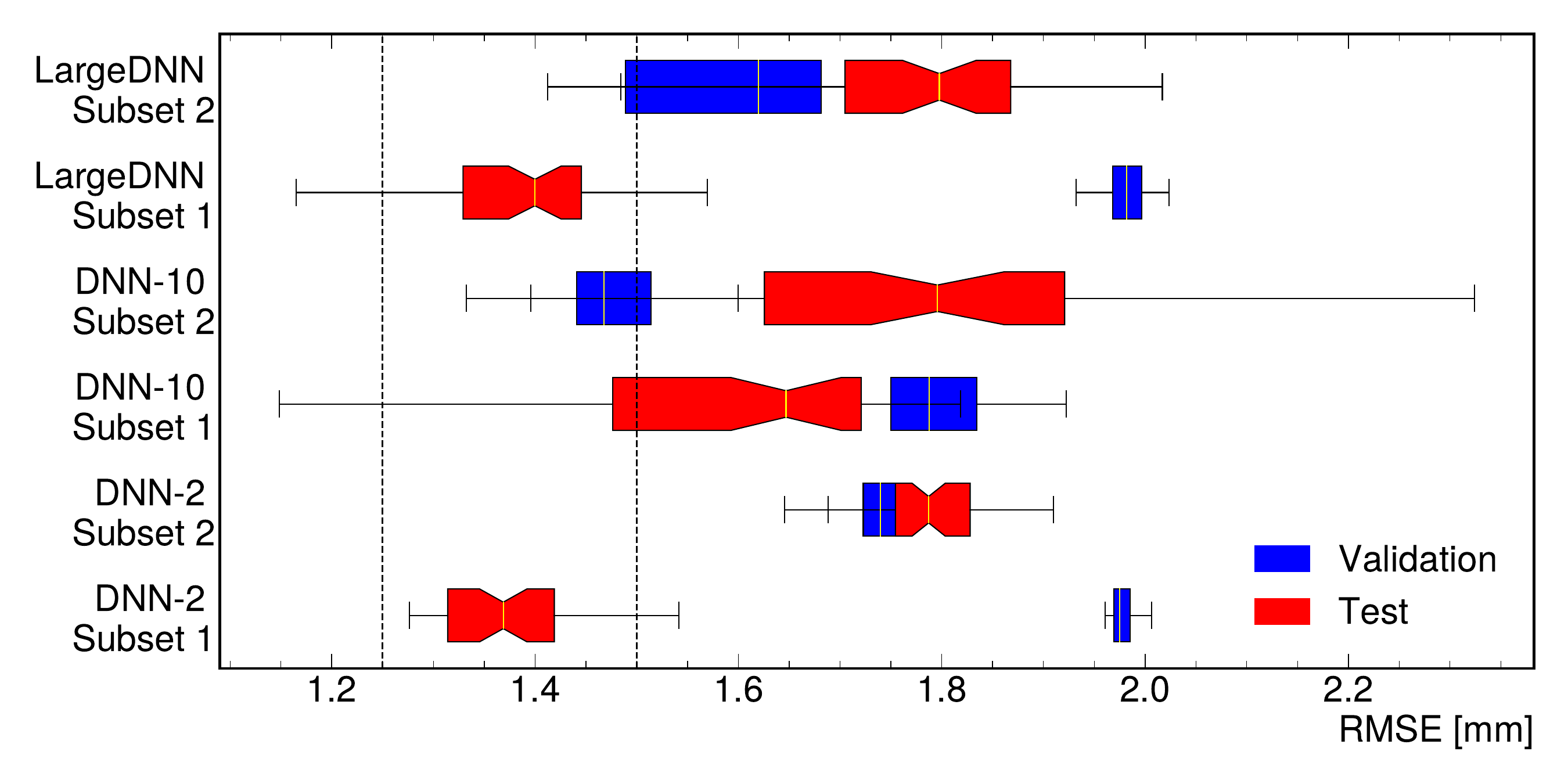}
    \end{center}
    \caption{Performance from selected models using subset 1 ($\text{C}_{\text{start}}$, $\text{A}_{\text{rise}}$, $\text{F}_{\text{start}}$) and subset 2 ($\text{B}_{\text{start}}$, $\text{C}_{\text{start}}$, $\text{D}_{\text{start}}$, $\text{F}_{\text{width}}$) of the reduced data set. All models shown trained with L1ES.}
    \label{fig:sr_subset_data_plot}
\end{figure}

Following the results of SR algorithms, we wished to see if neural networks could replicate similar performance for the same inputs. In~\cref{fig:sr_subset_data_plot} we display the performance of selected neural network models trained with L1ES, with the same input subsets from our SR studies. For brevity we will refer to the three inputs chosen by linear correlation ($\text{C}_{\text{start}}$, $\text{A}_{\text{rise}}$, $\text{F}_{\text{start}}$) and the four inputs chosen by the SR training ($\text{B}_{\text{start}}$, $\text{C}_{\text{start}}$, $\text{D}_{\text{start}}$, $\text{F}_{\text{width}}$) as subsets 1 and 2 respectively. We see that the neural networks typically outperform the SR algorithms, with subset 1 resulting in the lowest RMSE values on the HOS set.

Additionally, we observe an interesting trend where networks using subset 1 have larger discrepancies between the validation and HOS RMSE. While we have seen this in previous studies and labeled it as indicative of worse generalization, this is the first occurrence where the HOS RMSE is lower than the validation RMSE. By contrast the collection of inputs in subset 2 - those selected by the SR library - seem to show a greater alignment between the RMSE values for both the validation and HOS locations, even if the HOS RMSE values are higher than for subset 1. This holds true both for the output functions from the symbolic regression and neural network trainings, and serves as an interesting example of the viability for symbolic regression not just for algorithm compression but also for filtering features that are potentially detrimental to generalization.

\section{Remarks and Conclusions}
\label{sec:conclusion}


\begin{table}[!htpb]
    \caption{Summary of position reconstruction methods performed in this study and the best performing models and data sets for each method. Mean RMSE and standard deviations are reported for instances where we ran many different initializations of the model. Methods are ordered to match the chronological presentation in this paper.}
    \begin{center}
    \makebox[\textwidth]{%
    \begin{tabular}{lllc}
    \toprule
         Reconstruction Method & Best Model& Available Variables & Test (HOS) RMSE [mm]\\
         \midrule
         Single Parameter Fit & \hyperref[fig:Start]{$\sinh$ Fit} & $\text{C}_{\text{start}}$ & $\boldsymbol{2.125}$\\
         Classical Regression & \hyperref[fig:losssummary]{Ridge} & Full Dataset (X-nA) & $\boldsymbol{1.296\pm0.031}$\\
         Dropout Neural Networks & \hyperref[fig:DNNLossSummary]{DNN-2} & Full Dataset (X-nA) & $\boldsymbol{1.555\pm0.252}$\\
         L1ES Neural Networks & \hyperref[fig:DNNLossSummary2]{LargeDNN} & Reduced Dataset (S) & $\boldsymbol{1.440\pm0.120}$\\
         Symbolic Regression & \hyperref[eq:sr_four_param]{4-Param. Fit} & $\text{B}_{\text{start}}$, $\text{C}_{\text{start}}$, $\text{D}_{\text{start}}$, $\text{F}_{\text{width}}$ & $\boldsymbol{1.393}$\\
         \bottomrule
    \end{tabular}%
    }
    \end{center}
    \label{tab:conclusion_summary}
\end{table}

This paper describes the collection and attributes of the physical dataset obtained at the University of Minnesota using the prototype SuperCDMS detector. Studies presented here establish an initial benchmark for empirically studying the relationship between the location of the interaction and the physical pulse shapes observed at the different detector channels using methods of statistical and machine learning. We found that using the detailed timing information from the pulse shapes in different channels, the position of interaction can be reliably constructed using both linear regression and neural network models. Uncertainties associated with the reconstruction are found to be comparable with the experimental and detector resolution. It is valuable to note  that the timing parameters alone provided reasonable discrimination without using amplitude information.  In fact,  we observed that the inclusion of the amplitude information deteriorated the model's ability to generalize for previously unobserved impact locations. This is expected since the amplitude in this particular data set was poorly calibrated due to voltage drift.  

An unexpected result was that the simpler set of linear regression models provided superior and more robust performance in terms of reducing both reconstruction and generalization errors. The performance of DNNs varied significantly across different combinations of architecture and regularization choices. This counterintuitive observation can be partly attributed to the sparse coverage of the source data. Having the source interactions along a radial line effectively reduced the position reconstruction problem to a 1D problem, which did not uniformly populate channel shapes and edge types. 
The flexibility of DNN models to capture the intricate correlation structure of the pulse shapes probably would have benefitted from a larger dataset.
The degree to which a larger data set with better coverage improves the performance will yield information on ML strategies that should be followed as a function of the data quality and comprehensiveness. 
Our studies show that a combination of regularization and early stopping algorithms can be useful to counter the overfitting tendencies in larger neural networks. These observations will serve as a benchmark for future explorations on using these models in the context of  larger future datasets with two-dimensional scan of the physical interaction location.  
Building on the results and observations presented in this work, our future work will take a deeper look into overcoming the learning limitations of larger networks in the context of SuperCDMS datasets and explore  novel architectures like graph nets and transformers to investigate the full potential of modern machine learning the context of the SuperCDMS experiment.
\section{Acknowledgements}
This research is supported by the U.S. Department of Energy (DOE) Office of Science with award DE-SC0021395 from the Office of Advanced Scientific Computing Research (ASCR) by the FAIR Data Program of the DOE, Office of Science, ASCR, and by the US DOE Office of High Energy Physics under the award DE-SC0020185. The authors also acknowledge the Minnesota Supercomputing Institute (MSI) at the University of Minnesota for providing resources that contributed to the research results reported within this paper. 

\newpage  
\printbibliography

\end{document}